\documentclass[%
 aip,
 amsmath,amssymb,
 reprint,%
]{revtex4-1}

\usepackage{graphicx}% Include figure files
\usepackage{dcolumn}% Align table columns on decimal point
\usepackage{bm}% bold math
\usepackage[utf8]{inputenc}
\usepackage[T1]{fontenc}
\usepackage{mathptmx}
\usepackage{etoolbox}

\usepackage{color}

\makeatletter
\def\@email#1#2{%
 \endgroup
 \patchcmd{\titleblock@produce}
  {\frontmatter@RRAPformat}
  {\frontmatter@RRAPformat{\produce@RRAP{*#1\href{mailto:#2}{#2}}}\frontmatter@RRAPformat}
  {}{}
}%
\makeatother
\begin{document}

\preprint{AIP/123-QED}

\title[Polarization and magnetization in collisional and turbulent transport processes]{Polarization and magnetization in collisional and turbulent transport processes}
% Force line breaks with \\
\author{H. Sugama$^*$}
\email[Author to whom correspondence should be addressed:]{sugama.hideo@nifs.ac.jp}
%\thanks{}
\affiliation{
National Institute for Fusion Science, 
Toki 509-5292, Japan
}
\affiliation{
Department of Advanced Energy, University of Tokyo, 
Kashiwa 277-8561, Japan
}

\author{S. Matsuoka}
\affiliation{
National Institute for Fusion Science, 
Toki 509-5292, Japan
}
\affiliation{
Department of Fusion Science, SOKENDAI (The Graduate University for Advanced Studies), 
Toki 509-5292, Japan 
}

\author{M. Nunami}
\affiliation{
National Institute for Fusion Science, 
Toki 509-5292, Japan
}
\affiliation{
Department of Physics, Nagoya University, 
Nagoya 464-8602, Japan
}

\date{\today}% It is always \today, today,

\begin{abstract}
%===revision===
%\textcolor{red}{
Expressions of polarization and magnetization in magnetically confined plasmas 
are derived, which include full expansions in the gyroradius to treat effects of both equilibrium and 
microscopic electromagnetic turbulence.  
%}
%===
%
Using the obtained expressions, densities and flows of particles are related to 
those of gyrocenters. 
To the first order in the normalized gyroradius expansion, 
the mean part of the particle flow is given by the sum of the gyrocenter flow and the magnetization flow, which corresponds to the so-called magnetization 
law in drift kinetics,  
while the turbulent part contains the polarization flow as well.
Collisions make an additional contribution to the second-order particle flow. 
The mean particle flux across the magnetic surface is of 
the second-order and it contains classical, neoclassical, 
and turbulent transport processes.
%
%===revision===
%\textcolor{red}{
The Lagrangian variational principle is used to derive   
the gyrokinetic Poisson and Amp\`{e}re equations which properly 
include mean and turbulent parts so as to be useful for 
full-$f$ global electromagnetic gyrokinetic simulations. 
      It is found that the second-order Lagrangian term given 
by the inner product of the turbulent vector potential and the drift velocity consisting of 
the curvature drift and the $\nabla B$ drift should be retained 
in order for the derived Amp\`{e}re equation to correctly include the diamagnetic current 
which is necessary especially for the full-$f$ high-beta plasma simulations. 
%}
%===
%
The turbulent parts of these gyrokinetic Poisson and Amp\`{e}re equations 
are confirmed to agree with 
the results derived from the WKB representation in earlier works. 
\end{abstract}

\maketitle

\section{INTRODUCTION}

Global simulations of collisional and turbulent plasma transport~\cite{GTC,Idomura2017,XGC,Wang2009,GYSELA,ORB5,ELMFIRE,Gkeyll,Matsuoka} 
are now vigorously conducted based on gyrokinetic equations using the gyrocenter coordinates that are derived from the 
Lie transformation method.~\cite{B&H,Littlejohn1982} 
Conservation properties possessed by such gyrokinetic equations are suitable for global and long-time transport simulations and they have been extensively 
investigated based on Lagrangian and Hamiltonian 
formulations.~\cite{B&H,Sugama2000,Scott,Brizard2011,Parra_PPCF2011,
Sugama2017,Sugama2018,Sugama2021,Hirvijoki,Qin2014,Fan,Brizard2021}
It is well-known that the finite gyroradius representing 
the distance between particle and gyrocenter positions generates so-called polarization and magnetization,~\cite{B&H,Sugama2000} in terms of 
which the relations of the density and mean velocity of particles 
to those of gyrocenters are expressed. 
These relations are important for using gyrokinetic simulation results to correctly evaluate particle transport, as well as to accurately calculate the charge density and the electric current in Poisson and Amp\`{e}re equations, which are required to 
self-consistently determine electromagnetic fields in the simulation. 

In a general framework of macroscopic electromagnetism for material media 
consisting of molecules, 
polarization and magnetization are formulated for evaluating 
the macroscopic charge density and current by spatially averaging 
the microscopic density and current of the point charge around the 
center of mass of the molecule.~\cite{Jackson} 
Then, the resultant expressions of the macroscopic charge density and current 
are given by the series expansion associated with multipole moments 
due to the finite distance of each point charge from the center of mass of 
the molecule. 
The local spatial average and the finite distance described above   
for the system of molecules are replaced by the phase-space integral 
including the distribution function 
and the finite gyroradius of the particle motion around the gyrocenter, 
respectively, for 
formulating the polarization and magnetization 
in the gyrokinetic system considered in the present study.  
In the drift kinetic system without microturbulence, 
the particle flow is represented by the sum of the gyrocenter flow and 
the magnetization flow, which is called the magnetization 
law.~\cite{magnetization_law} 
In this work, we use the gyrocenter phase-space coordinates obtained from the 
particle phase-space coordinates using the Lie transformation, by which 
the effects of turbulent electromagnetic fields are included in definitions 
of the gyrocenter position and the gyroradius vector. 
This gyroradius vector is used for infinite series expansion to express the 
polarization and magnetization in magnetically confined 
plasmas with gyroradius scale fluctuations. 

The polarization and magnetization are also derived from taking the variational 
derivative of the field-particle interaction part of 
the Lagrangian for the system with respect to the electric and 
magnetic fields, respectively.~\cite{B&H} 
This derivation is not commonly used in conventional gyrokinetic 
studies~\cite{Antonsen,CTB,F-C} 
where scalar and vector potentials are used instead of electromagnetic fields 
to formulate basic equations describing plasma microturbulence.  
In some recent studies,~\cite{Burby2019,Brizard2021}
the Lagrangian of the gyrokinetic system is expressed in terms of
perturbed electromagnetic fields instead of 
perturbed scalar and vector potentials, so that 
the gyrokinetic polarization and magnetization 
can be obtained by the derivative of the Lagrangian. 
The scalar and vector potentials are used in our study where  
conventional studies' results on gyrokinetic Poisson and 
Amp\`{e}re equations with the polarization and magnetization effects 
due to turbulent fields are consistently incorporated. 
In addition, the magnetization law in drift kinetics is reproduced 
from taking the ensemble average of 
the expression for the particle flow obtained in this paper. 
To the second order in the normalized gyroradius, 
the effect of the collision term, which is not described in the magnetization 
law, appears as the classical transport~\cite{Hinton1976,H&S,Helander}
 in the ensemble-averaged particle 
flow. 
Then it is confirmed that in toroidal confinement confinement systems, 
the average particle flux across  
the magnetic surface is given by the second-order flows in which 
the classical, neoclassical,~\cite{Hinton1976,H&S,Helander} 
and turbulent transport~\cite{Horton} are included. 

The rest of this paper is organized as follows.
In Sec.~II,
the densities and flows of the particles and gyrocenters are defined
using velocity-space integrals of the distribution
functions in the particle and gyrocenter phase-space coordinates.
Then, the gyrocenter and particle transport equations derived from 
the Boltzmann kinetic equations in the two coordinate systems 
are used to obtain the relation between
the particle and gyrocenter flows, in which effects of polarization,
magnetization, and collisions are included.
The detailed expressions of the polarization and magnetization are 
presented in Sec.~III. 
In Sec.~IV, the particle flows due to gyrocenter motion, polarization, 
magnetization, and collisions are separately treated using expansion in the normalized 
gyroradius parameter and decomposition into the ensemble average and turbulent parts.
There it is shown that 
the first-order  ensemble-averaged particle flow obeys
the so-called magnetization law in drift kinetics, while 
the mean particle flux across the magnetic surface is of 
the second order and contains classical, neoclassical, 
and turbulent transport processes. 
The Lagrangian for variational derivation of the gyrokinetic 
Vlasov equation, Poisson's equation, and Amp\`{e}re's law is 
presented in Sec.~V, where the linear polarization-magnetization 
approximation~\cite{Sugama2000} is employed. 
Finally, conclusions are given in Sec.~VI. 
In addition, Appendix~A presents the transformation formulas from 
the particle coordinates to the gyrocenter coordinates, and 
the gyrocenter Lagrangian, from which the gyrocenter equations of 
motion are derived. 
In Appendix~B, the gyrocenter velocity and the time derivative of 
the gyroradius vector are expanded in the normalized 
gyroradius parameter to obtain useful formulas for derivation of 
the results given in Sec.~IV. 
The zeroth and first-order distribution functions 
and the conditions satisfied by them are described in Appendix~C. 
It is verified in Appendix~D that 
the turbulent parts of Poisson and  Amp\`{e}re equations obtained 
in the present work agree with the results derived in 
earlier works
using the WKB representation.~\cite{Antonsen,CTB}

\section{DENSITIES AND FLOWS OF PARTICLES AND GYROCENTERS}

The gyrokinetic Boltzmann equation for 
the gyrocenter distribution function 
$f_a({\bf Z}, t)$ of the particle species $a$ is given by 
\begin{equation}
\label{2-1}
\frac{df_a}{dt} \equiv  
\left(
\frac{\partial}{\partial t}
+ 
\frac{d {\bf Z}}{dt} \cdot 
\frac{\partial}{\partial {\bf Z}}
\right) f_a
 = C_a^{(g)} 
,
\end{equation}
where the gyrocenter phase-space coordinates 
$
{\bf Z} \equiv 
({\bf X}, U, \mu, \xi)
$
are defined in terms of the particle phase-space coordinates 
$
{\bf z} \equiv 
({\bf x}, v_\parallel, \mu_0, \xi_0)
$
as shown in Appendix~A [see Eqs.~(\ref{A-6})--(\ref{A-9})]. 
In Eq.~(\ref{2-1}), $d {\bf Z}/dt$ is regarded as a function of 
$({\bf Z}, t)$, which is given by
the gyrocenter motion equations, 
Eqs~(\ref{A-27})--(\ref{A-31}).
The collision term $C_a^{(g)}$ in the gyrocenter coordinates 
is written as~\cite{Brizard2004,Sugama2015} 
\begin{equation}
\label{2-2}
 C_a^{(g)}  
 \equiv
\sum_b
C_{ab}^{(g)} [f_a, f_b]
\equiv 
\sum_b 
{\cal T}_a^{-1*}C_{ab}^{(p)} 
[{\cal T}_a^* f_a, {\cal T}_b^* f_b]
,
\end{equation}
where the subscripts $a$ and $b$ represent 
species of colliding particles and   
$C_{ab}^{(p)}[f_a^{(p)}, f_b^{(p)}]$ is 
the Landau collision operator~\cite{Hinton1976,H&S,Helander} 
for the distribution functions 
$f_a^{(p)} = {\cal T}_a^* f_a$ and 
$f_b^{(p)} = {\cal T}_b^* f_b$ in the particle coordinates, 
which are 
obtained by the pull-back operators 
${\cal T}_a^*$ and ${\cal T}_b^*$ acting on 
the gyrocenter distribution functions 
$f_a$ and $f_b$, respectively. 
It is noted that, for the function $f_a ({\bf Z})$ of 
the gyrocenter coordinates 
${\bf Z}$, ${\cal T}_a^* f_a$ is defined by 
$({\cal T}_a^* f_a) ({\bf z}) \equiv f_a ( {\cal T}_a ({\bf z}) )$, 
where ${\bf Z} = {\cal T}_a ({\bf z})$ represents the transformation from 
the particle coordinates to the gyrocenter coordinates. 
The detailed expressions of the coordinate transformation 
are shown in Appendix~A. 
The push-forward operator ${\cal T}_a^{-1*}$ 
is used to obtain the expression of the collision 
term in the gyrocenter coordinates from 
that in the particle coordinates. 
For the function $C_a^{(p)} ({\bf z})$ of the particle coordinates 
${\bf z}$, 
${\cal T}_a^{-1*} C_a^{(p)}$ is defined by 
$({\cal T}_a^{-1*} C_a^{(p)})({\bf Z}) \equiv 
C_a^{(p)} ({\cal T}_a^{-1} ({\bf Z}) )$
where ${\bf z} = {\cal T}_a^{-1} ({\bf Z})$ represents 
the transformation from 
the gyrocenter coordinates to the particle coordinates.

The gyrophase average of an arbitrary function $Q$ of the gyrocenter 
phase-space coordinates 
${\bf Z} \equiv ({\bf X}, U, \mu, \xi)$ is represented by  
\begin{equation} 
\label{2-3}
\langle Q \rangle_\xi 
\equiv
\frac{1}{2\pi}
\oint  Q \, d\xi
, 
\end{equation}
and the gyrophase-dependent part of $Q$ is written as
\begin{equation}
\label{2-4}
\widetilde{Q}
\equiv
Q -
\langle Q \rangle_\xi
.
\end{equation}
From Eq.~(\ref{2-1}), we obtain 
\begin{equation}
\label{2-5}
\frac{d\langle f_a\rangle_\xi}{dt} \equiv  
\left(
\frac{\partial}{\partial t}
+ 
\frac{d {\bf Z}}{dt} \cdot 
\frac{\partial}{\partial {\bf Z}}
\right) 
\langle f_a\rangle_\xi
 = \langle C_a^{(g)} \rangle_\xi
,
\end{equation}
and
\begin{equation}
\label{2-6}
\frac{d\widetilde{f}_a}{dt} \equiv  
\left(
\frac{\partial}{\partial t}
+ 
\frac{d {\bf Z}}{dt} \cdot 
\frac{\partial}{\partial {\bf Z}}
\right) \widetilde{f}_a
 = \widetilde{C}_a^{(g)} 
.
\end{equation}
Noting that the pull-back ${\cal T}_a^* f_a$ included in  
the definition of the gyrocenter collision operator 
$C^{(g)}$ 
has a gyrophase dependence different from what $f_a$ has, 
we find that the gyrocenter collision term depends on the gyrophase 
angle $\xi$ even when the operator $C_a^{(g)}$ acts on the 
gyrocenter distributions which are independent of $\xi$. 
Therefore, $\widetilde{C}_a^{(g)}$ does not vanish generally, and 
Eq.~(\ref{2-6}) yields the nonzero gyrophase-dependent part 
$\widetilde{f}$ of the gyrocenter distribution function. 
Using the gyrofrequency $\Omega_a \equiv e_a B/(m_a c)$ to 
approximately write 
$d\widetilde{f}_a/dt \simeq \Omega_a \partial\widetilde{f}_a/\partial \xi$, 
we have
\begin{equation}
\label{2-7}
\widetilde{f}_a 
\simeq
 \frac{1}{\Omega_a} \int^\xi \widetilde{C}_a^{(g)} d\xi
= {\cal O} (\epsilon^2 f_{a0} )
,  
\end{equation}
where $\widetilde{C}_a^{(g)}[f_a] \simeq \widetilde{C}_a^{(g)}[f_{a1}]
= {\cal O}(\nu_a \epsilon f_{a0})$ 
and  
$\nu_a / \Omega_a = \epsilon \, \nu_a / \omega_{Ta} = 
{\cal O}(\epsilon)$ 
are used. 
Here, 
$f_{a0}$ and $f_{a1}$ are the zeroth and first-order 
distribution functions [see Eq.~(\ref{4-1})]
in the expansion with respect to the normalized gyroradius 
parameter $\epsilon$  
given by the ratio of the gyroradius 
$\rho_a$ to the equilibrium scale length $L$. 
As for the ratio $\nu_a / \omega_{Ta}$ of the collision frequency $\nu_a$ 
to the transit frequency $\omega_{Ta} \sim L/v_{Ta}$ 
[$v_{Ta} \equiv (2T_a/m_a)^{1/2}$: the thermal velocity], 
we do not consider a subsidiary ordering 
such as those used in the Pfirsh-Schl\"{u}ter, plateau, 
and banana regimes.~\cite{Hinton1976,H&S,Helander}

The Boltzmann equation for the distribution function 
$f_a^{(p)}$ of the particle species $a$
in the particle coordinates 
${\bf z} \equiv ({\bf x}, v_\parallel, \mu_0, \xi_0)$
is written as 
\begin{equation}
\label{2-8}
\frac{df_a^{(p)}}{dt} \equiv  
\left(
\frac{\partial}{\partial t}
+ 
\frac{d {\bf z}}{dt} \cdot 
\frac{\partial}{\partial {\bf z}}
\right) f_a^{(p)}
 = C_a^{(p)} 
.
\end{equation}
The particle density $n_a^{(p)}$ and 
the particle flow 
$\boldsymbol{\Gamma}_a^{(p)}$  
are defined as functions of 
the position ${\bf x}$ and the time $t$ by
\begin{equation}
\label{2-9}
n_a^{(p)} ({\bf x}, t) 
=
\int d^6 z'
\; \delta^3 ({\bf x}' - {\bf x} ) 
D_a^{(p)} ({\bf x}', t) f_a^{(p)} ({\bf z}', t) 
,
\end{equation}
and
\begin{equation}
\label{2-10}
\boldsymbol{\Gamma}_a^{(p)}
 ({\bf x}, t) 
=
\int d^6 z'
\;  
\delta^3 ({\bf x}' - {\bf x} )
D_a^{(p)} ({\bf x}', t) f_a^{(p)} ({\bf z}', t) 
{\bf v}'
,
\end{equation}
respectively,
where the Jacobian 
$D_a^{(p)} ({\bf x}, t) \equiv B({\bf x}, t)/m_a$ is used. 

Multiplying Eq.~(\ref{2-8}) by 
$D_a^{(p)}$ and integrating it with respect to 
the velocity space variables 
$v_\parallel$, $\mu_0 \equiv m_a v_\perp^2/ (2B)$, and $\xi_0$, 
we obtain the continuity equation
\begin{equation}
\label{2-11}
\frac{\partial n_a^{(p)}({\bf x}, t)}{\partial t} +  
\nabla \cdot \boldsymbol{\Gamma}_a^{(p)} ({\bf x}, t)
 = 0
,
\end{equation}
where the particle number conservation in collisions, 
$\int d^6 z' \, \delta ({\bf x} - {\bf x}') D_a^{(p)} ({\bf x}', t) 
C_a^{p} ({\bf z}', t) = 0$,  
is used. 
Similarly, multiplying Eq.~(\ref{2-1}) by 
$D_a ({\bf Z}, t) \equiv B^*_{a\parallel} ({\bf Z}, t)/m_a$ 
[see Eq.~(\ref{A-32}) for the definition of $B^*_{a\parallel}$]
and integrating it with respect to 
the velocity space variables 
$U$, $\mu$, and $\xi$, 
we obtain 
\begin{eqnarray}
\label{2-12}
& & \frac{\partial n_a^{(g)}({\bf x}, t)}{\partial t} +  
\nabla \cdot \boldsymbol{\Gamma}_a^{(g)} ({\bf x}, t)
\nonumber \\ & & \mbox{}
 = 
\int d^6 Z' \; 
\delta^3 ({\bf X}' - {\bf x} ) 
D_a ({\bf Z}', t) C_a^{(g)} ({\bf Z}', t)
\nonumber \\ & & \mbox{}
= 
- \nabla \cdot \boldsymbol{\Gamma}_a^{C} 
({\bf x}, t)
\end{eqnarray}
where
the gyrocenter density $n_a^{(g)}$ and 
the gyrocenter flow 
$\boldsymbol{\Gamma}_a^{(g)}   
 \equiv  n_a^{(g)} {\bf u}_a^{(g)}$
are defined by
\begin{equation}
\label{2-13}
n_a^{(g)} ({\bf x}, t) 
=
\int d^6 Z
\; D_a ({\bf Z}, t) f_a ({\bf Z}, t) 
\delta^3 ({\bf X} - {\bf x} )
,
\end{equation}
and
\begin{eqnarray}
\label{2-14}
 \boldsymbol{\Gamma}_a^{(g)} ({\bf x}, t)  
&  \equiv  &
n_a^{(g)} {\bf u}_a^{(g)} ({\bf x}, t) 
\nonumber \\ 
& \equiv  & 
\int d^6 Z
\; D_a ({\bf Z}, t) f_a ({\bf Z}, t) 
\delta^3 ( {\bf X} - {\bf x} )
\frac{d {\bf X}}{dt} 
, 
\hspace*{3mm}
\end{eqnarray}
respectively. 
The gyrocenter velocity $d {\bf X}/dt$ which enters 
the integrand in Eq.~(\ref{2-14}) 
is regarded as a function of $({\bf Z}, t)$ using 
Eq.~(\ref{A-28}). 
As shown in Ref.~\cite{Sugama2015}, 
$\boldsymbol{\Gamma}_a^C$ on the right-hand side of 
Eq.~(\ref{2-12})
is given by 
\begin{eqnarray}
\label{2-15}
\boldsymbol{\Gamma}_a^C ({\bf x}, t)  
&  \equiv  & 
\sum_{l=0}^\infty 
\frac{(-1)^{l}}{(l+1)!} \frac{\partial^l}{\partial x^{j_1}
\cdots \partial x^{j_l}}
\biggl(
\int d^6 z'
\;
\delta^3 ( {\bf x}' - {\bf x} )
\nonumber \\ 
&  &
\cdot
\;
 D_a^{(p)} 
\; \sum_b C_{ab}^{(p)}[{\cal T}_a^* f_a, {\cal T}_b^* f_b] 
\;
 \Delta {\bf x}_a
\; \Delta x_a^{j_1} \cdots \Delta x_a^{j_l}
\biggr)
, 
\nonumber \\ & & 
\end{eqnarray}
where $\Delta {\bf x}_a \equiv {\bf X} - {\bf x}$ is defined 
as a function of ${\bf z}$ using Eq.~(\ref{A-6})  
and  
$\Delta x_a^j$ is its $j$th component. 
As seen later in Eq.~(\ref{4C-5}),  
the classical particle transport is derived from  
$\boldsymbol{\Gamma}_a^C$.

The particle density $n_a^{(p)}$ and 
the gyrocenter density $n_a^{(g)}$ are related to each other 
by
\begin{equation}
\label{2-16}
e_a \, n_a^{(p)}  
=
e_a \, n_a^{(g)} 
- \nabla \cdot {\bf P}_a, 
\end{equation}
where ${\bf P}_a$ is the polarization vector due to the particle species $a$,  
and its detailed expression is presented later in Eq.~(\ref{3-2}). 
The polarization current due to the 
particle species $a$ is given by
\begin{equation}
\label{2-17}
{\bf J}_a^{\rm pol} 
\equiv
e_a \,
\boldsymbol{\Gamma}_a^{\rm pol} 
\equiv 
 \frac{\partial {\bf P}_a}{\partial t}
,
\end{equation}
where $\boldsymbol{\Gamma}_a^{\rm pol}$ represents 
the polarization particle flow of the species $a$. 
 
It is shown in Sec.~III that 
the particle flow $\boldsymbol{\Gamma}_a^{(p)}$ 
is written as  
\begin{equation}
\label{2-18}
\boldsymbol{\Gamma}_a^{(p)} 
=
\boldsymbol{\Gamma}_a^{(g)} 
+
\boldsymbol{\Gamma}_a^{\rm pol} 
+
\boldsymbol{\Gamma}_a^{\rm mag} 
+
\boldsymbol{\Gamma}_a^{C*}
,
\end{equation}
where $\boldsymbol{\Gamma}_a^{C*}$ is defined 
later in Eq.~(\ref{2-15*}) and it satisfies 
$
\nabla \cdot \boldsymbol{\Gamma}_a^{C*} 
=
\nabla \cdot \boldsymbol{\Gamma}_a^C
$. 
Here, $\boldsymbol{\Gamma}_a^{\rm mag}$ represents  
the particle flow due to the magnetization 
which is defined by
\begin{equation}
\label{2-19}
{\bf J}_a^{\rm mag} 
\equiv
e_a \,
\boldsymbol{\Gamma}_a^{\rm mag} 
\equiv 
c \,
\nabla \times {\bf M}_a 
\end{equation}
where ${\bf M}_a$ and ${\bf J}_a^{\rm mag}$ are 
the magnetization vector and the magnetization current density due to 
the particle species $a$, respectively. 
The detailed expression of ${\bf M}_a$ is shown later in Eq.~(\ref{3-7}). 
Subtracting Eq.~(\ref{2-12}) from Eq.~(\ref{2-11}),  
we have 
\begin{equation}
\label{2-20}
\frac{\partial (n_a^{(p)} - n_a^{(g)})}{\partial t} +  
\nabla \cdot 
(
\boldsymbol{\Gamma}_a^{(p)} 
- \boldsymbol{\Gamma}_a^{(g)} 
)
= 
 \nabla \cdot \boldsymbol{\Gamma}_a^{C} 
\end{equation}
We can easily verify that Eq.~(\ref{2-20}) 
is satisfied by Eqs.~(\ref{2-16})--(\ref{2-19}).

\section{POLARIZATION AND MAGNETIZATION FLOWS}

Performing the transformation from the particle coordinates to the gyrocenter coordinates for the integration in Eq.~(\ref{2-9}), 
we obtain
\begin{eqnarray}
\label{3-1}
n_a^{(p)} ({\bf x}, t) 
& = & 
\int d^6 Z
\; D_a ({\bf Z}, t) f_a ({\bf Z}, t) 
\delta^3 [{\bf X} + 
\boldsymbol{\rho}_a({\bf Z}, t) - {\bf x} ]
\nonumber \\ 
& = & 
n_a^{(g)} ({\bf x}, t) 
- \nabla \cdot [ e_a^{-1} {\bf P}_a({\bf x}, t) ]
,
\end{eqnarray}
where the gyroradius vector $\boldsymbol{\rho}_a$ 
is defined by Eqs.~(\ref{A-10})--(\ref{A-17}) in Appendix~A and 
the polarization vector ${\bf P}_a({\bf x}, t)$ 
is given by
\begin{eqnarray}
\label{3-2}
\frac{1}{e_a}{\bf P}_a({\bf x}, t)
& = & 
\sum_{l=0}^\infty 
\frac{(-1)^l}{(l+1)!} \frac{\partial^l}{\partial x^{j_1}\cdots 
\partial x^{j_l}}
\biggl(
\int d^6 Z
\; \delta^3 ({\bf X} - {\bf x} )
\nonumber \\ & & 
\times D_a ({\bf Z}, t) f_a ({\bf Z}, t) 
\; \boldsymbol{\rho}_a
\; \rho_a^{j_1} \cdots \rho_a^{j_l}
\biggr)
.
\end{eqnarray}
The $j$th components of the vectors 
${\bf x}$ and $\boldsymbol{\rho}_a$ 
are denoted by $x^j$ and $\rho_a^j$, respectively.  
Here and hereafter, 
we employ the summation convention 
that the same symbol used for a pair of indices in 
upper and lower positions within a term 
[such as in Eq.~(\ref{3-2})] 
indicates summation over the range $\{1, 2, 3\}$ of the symbol index.  
In deriving Eqs.(\ref{3-1}) and (\ref{3-2}), the Taylor expansion,
\begin{equation}
\label{3-3}
\delta^3 ({\bf X} + 
\boldsymbol{\rho}_a - {\bf x})
=
\sum_{l=0}^\infty
\frac{(-1)^l}{l!}
\rho_a^{j_1} \cdots \rho_a^{j_l}
 \frac{\partial^l \delta^3 ({\bf X}  - {\bf x})}{\partial x^{j_1}
\cdots \partial x^{j_l}}
,
\end{equation}
is used and partial integrations are 
performed. 
Taking the partial time derivative of Eq.~(\ref{3-2}) and 
using Eq.~(\ref{2-1}), 
we find that the polarization flow 
$\boldsymbol{\Gamma}_a^{\rm pol}$ due to the 
particle species $a$ is given by 
\begin{eqnarray}
\label{3-4}
& & 
\boldsymbol{\Gamma}_a^{\rm pol} 
\equiv 
\frac{1}{e_a}\frac{\partial {\bf P}_a({\bf x}, t)}{\partial t}
\nonumber \\ 
& & =
\sum_{l=0}^\infty 
\frac{(-1)^l}{(l+1)!} \frac{\partial^l}{\partial x^{j_1}
\cdots \partial x^{j_l}}
\biggl[
\int d^6 Z  \; 
\delta^3 ({\bf X} - {\bf x} )
\nonumber \\ & & 
\hspace*{5mm} 
\times
\biggl\{
D_a f_a
\biggl(
\frac{d \boldsymbol{\rho}_a}{dt}
 \rho_a^{j_1} \cdots  \rho_a^{j_l}
+ l \; \boldsymbol{\rho}_a
\; \frac{d \rho_a^{j_1}}{dt} 
\rho_a^{j_2} \cdots \rho_a^{j_l}
\biggr)
\nonumber \\ & &  \mbox{}
\hspace*{6mm}
- \frac{\partial}{\partial {\bf X}} \cdot
\biggl(
D_a f_a \frac{d {\bf X}}{dt}
 \boldsymbol{\rho}_a
\; \rho_a^{j_1} \cdots \rho_a^{j_l}
\biggr) \;
\biggr\} \:
\biggr]
- \boldsymbol{\Gamma}_a^{C*}
,
\hspace*{5mm}
\end{eqnarray}
where $\boldsymbol{\Gamma}_a^{C*}$ is defined by
\begin{eqnarray}
\label{2-15*}
\boldsymbol{\Gamma}_a^{C*} ({\bf x}, t)  
&  \equiv  & 
\sum_{l=0}^\infty 
\frac{(-1)^{l+1}}{(l+1)!} \frac{\partial^l}{\partial x^{j_1}
\cdots \partial x^{j_l}}
\biggl(
\int d^6 Z'
\;
\delta^3 ( {\bf X}' - {\bf x} )
\nonumber \\ 
&  &
\cdot
\;
 D_a 
\; \sum_b C_{ab}^{(g)}[f_a, f_b] 
\;
 \boldsymbol{\rho}_a
\; \rho_a^{j_1} \cdots \rho_a^{j_l}
\biggr)
. 
\end{eqnarray}
It can be shown from Eqs.~(\ref{2-15}) and (\ref{2-15*})
that 
$
\nabla \cdot \boldsymbol{\Gamma}_a^{C*} 
-\nabla \cdot \boldsymbol{\Gamma}_a^C
= \int d^6 z' \delta^3 ({\bf x}' - {\bf x}) D_a^{(p)} C_a^{(p)}
= 0
$ 
and accordingly 
$
\nabla \cdot \boldsymbol{\Gamma}_a^{C*} 
=
\nabla \cdot \boldsymbol{\Gamma}_a^C
$. 
In addition, as seen in Sec.~IV, 
both $\boldsymbol{\Gamma}_a^{C*}$  
and 
$\boldsymbol{\Gamma}_a^C$ are of 
${\cal O}(\epsilon^2)$ and 
their ensemble averages coincide with each other and 
represent the classical particle transport. 

The particle flow $\boldsymbol{\Gamma}_a^{(p)}$ 
defined in Eq.~(\ref{2-10}) is also given by 
the integration in the gyrocenter coordinates as 
\begin{eqnarray}
\label{3-5}
& & \boldsymbol{\Gamma}_a^{(p)} ({\bf x}, t) 
 \equiv 
n_a^{(p)} {\bf u}_a^{(p)} ({\bf x}, t) 
\nonumber \\ 
&  \equiv  & 
\int d^6 Z
\; D_a ({\bf Z}, t) f_a ({\bf Z}, t) 
\delta^3 ( {\bf X} + 
\boldsymbol{\rho}_a - {\bf x} )
\left(
\frac{d {\bf X}}{dt} 
+ \frac{d \boldsymbol{\rho}_a}{dt} 
\right)
.
\nonumber \\ & & 
\end{eqnarray}
where the particle velocity is represented by 
$d {\bf X} / dt
+ d \boldsymbol{\rho}_a/dt$,  
which is regarded as a function of $({\bf Z}, t)$,  
using Eq.~(\ref{A-28}) in Appendix~A and 
Eqs.~(\ref{B-6})--(\ref{B-9}) in Appendix~B. 
Then we can use Eqs.~(\ref{2-14}), (\ref{2-15}), and 
(\ref{3-2})--(\ref{3-4}) to derive 
Eq.~(\ref{2-18}) which 
is written here as 
\begin{eqnarray}
\label{3-6}
\boldsymbol{\Gamma}_a^{(p)} ({\bf x}, t) 
& = & 
\boldsymbol{\Gamma}_a^{(g)} ({\bf x}, t) 
+
\frac{1}{e_a}\frac{\partial {\bf P}_a({\bf x}, t)}{\partial t}
+
\frac{c}{e_a}\nabla \times {\bf M}_a ({\bf x}, t)
\nonumber \\ 
&  & 
\mbox{}
+ \boldsymbol{\Gamma}_a^{C*} ({\bf x}, t) 
,
\end{eqnarray}
where 
$\boldsymbol{\Gamma}_a^{\rm mag} \equiv 
(c/e_a)\nabla \times {\bf M}_a ({\bf x}, t)$ 
is the particle flow
due to the magnetization vector ${\bf M}_a$ defined by 
\begin{eqnarray}
\label{3-7}
& & \frac{c}{e_a}{\bf M}_a ({\bf x}, t) 
\nonumber \\ 
& \equiv  &
\sum_{l=0}^\infty 
\frac{(-1)^l}{l!} \frac{\partial^l}{\partial x^{j_1}
\cdots \partial x^{j_l}}
\biggl[
\int d^6 Z  \;  D_a f_a
\delta^3 ({\bf X} - {\bf x} )
\nonumber \\ & & \mbox{}
\cdot 
\rho_a^{j_1} \cdots \rho_a^{j_l}
\boldsymbol{\rho}_a \times
\biggl(
\frac{1}{(l+2)} 
\frac{d \boldsymbol{\rho}_a}{dt}
+
\frac{1}{(l+1)} 
\frac{d {\bf X}}{dt}
\biggr)
\biggr]
. 
\hspace*{5mm}
\end{eqnarray}

\section{EXPANSION OF PARTICLE FLOWS IN THE NORMALIZED GYRORADIUS PARAMETER $\epsilon$}

We here first expand the gyrocenter distribution function in 
the normalized gyroradius parameter $\epsilon$ as
\begin{equation}
\label{4-1}
f_a ({\bf Z}, t) = 
f_{a0} ({\bf Z}, t) + f_{a1} ({\bf Z}, t) + f_{a2} ({\bf Z}, t) + \cdots
,
\end{equation}
where the subscripts $n=0,1,2,\cdots$ represent 
the terms of ${\cal O}(\epsilon^n)$. 
More precisely speaking, 
$f_{an} = {\cal O}(\epsilon^n)$ implies that 
the magnitude of $f_n$ is represented by $f_{an} = {\cal O}( \epsilon^n f_{a0} )$.  
 
The gyrocenter distribution function is also written as the sum of 
the ensemble average part and the fluctuation part, 
\begin{equation}
\label{4-2}
f_a =  \langle f_a \rangle_{\rm ens} + \widehat{f}_a
. 
\end{equation}
We denote the average and fluctuation parts of the magnetic field by    
${\bf B} = \nabla \times {\bf A}$ and 
$\widehat{\bf B} = \nabla \times \widehat{\bf A}$, respectively. 
%
%===revision===
%\textcolor{red}{
The ensemble average is used 
as the basic method of statistical mechanics to obtain the macroscopic mean values of physical valuables. 
For the case of gyrokinetic turbulence simulation, 
an ensemble literally corresponds to a group of a large number of simulations performed 
using many different sets of randomly given initial perturbations 
while being done 
for the same macroscopic
state (characterized by the same conditions for background  profiles of 
densities, temperatures, and electromagnetic fields), 
and the ensemble average of some variable is defined by the average of its values obtained from 
the repeatedly performed simulations. 
However, assuming that 
a single typical nonlinear gyrokinetic simulation shows ergodic behavior wandering among 
a large number of microscopic turbulent states which will be
produced by the ensemble of simulations,  
the ensemble average is considered to equal 
the local space-time average obtained in the single simulation. 
This local space-time averaging of the distribution and other field functions 
in gyrokinetic systems is detailedly described in 
Ref.~\cite{Abel}, which shows the same results as given in Ref.~\cite{Sugama1998} using 
the notation of the ensemble average. 
%}
%===

%===revision===
%\textcolor{red}{
We note here that the gyrophase average should be clearly distinguished from 
the local space average related to the ensemble average. 
The ensemble average can be replaced by the space-time average over scales 
which are much smaller than macroscopic scales 
but sufficiently larger than microscopic fluctuation scales.~\cite{Abel} 
For example,  for the fluctuation potential 
$
\phi({\bf x}) = \phi_{{\bf k}_\perp} \exp ( i {\bf k}_\perp \cdot {\bf x} )
$
with the perpendicular wavenumber vector ${\bf k}_\perp$ $(k_\perp \sim \rho^{-1})$, 
the local space average  of  $\phi({\bf x})$ over the scale $l$ $(\rho \ll l \ll L)$ in the plane perpendicular to 
the background magnetic field vanishes. 
On the other hand, the gyrophase average 
$\langle \cdots \rangle_\xi$
of the fluctuating potential
is given by 
$
\langle \phi ({\bf X} + \boldsymbol{\rho} ) \rangle_\xi = 
J_0 (k_\perp \rho)  \phi_{\bf k} \exp ( i {\bf k}_\perp \cdot {\bf X} )
$
which shows that the gyrophase average does not completely remove the fluctuation but 
weakens it by the factor 
$J_0 (k_\perp \rho) =
\langle 
\exp ( i {\bf k}_\perp \cdot \boldsymbol{\rho} )
\rangle_\xi$
[which is derived from the formula, 
$
(2\pi)^{-1}\oint  \exp (i x \sin \theta )  d\theta = J_0 (x)
$]. 
%}
%===

%===revision===
%\textcolor{red}{
As seen in Eq.~(\ref{4-2}), the fluctuation part of the distribution function 
is  given as the deviation from the ensemble average. 
We now recall that, in the present work using the modern gyrokinetic formulation, 
the gyrocenter coordinates ${\bf Z}$ in 
$f_a = f_a ({\bf Z}, t)$ are defined from the particle coordinates ${\bf z}$
with effects of the electromagnetic fluctuations taken into account 
[see Eqs.~(\ref{A-6})--(\ref{A-9})]. 
On the other hand, in the classical gyrokinetic 
formulation~\cite{Antonsen,CTB,F-C}
using the WKB  representation (see Appendix~C) for the fluctuating parts of 
the distribution function and electromagnetic fields, 
the particle phase-space coordinates used as independent variables of the distribution 
function are defined without including effects of the fluctuations.  
Then, due to 
the difference between the two sets of the phase-space coordinates,  
the fluctuation part of the distribution function in the modern gyrokinetic formulation 
differs from that in the classical formulation 
[see Eq.~(\ref{C2-14}) in Appendix~C 
where $\widehat{f}_{a1}^{(p)}$ and $\widehat{f}_{a1}$ correspond to
the fluctuation parts of the distribution functions 
in the classical and modern formulations, respectively]. 
%}
%===

%===revision===
%\textcolor{red}{
In the rest of this section, 
the expansion in $\epsilon$ [Eq.~(\ref{4-1})] and 
the decomposition into the average and fluctuation parts 
[Eq.~(\ref{4-2})] 
are employed to analyze various components which compose the particle flow 
[Eq.~(\ref{3-6})]. 
It is noted that, even in the case without microscopic fluctuations, 
the expansion of the distribution function 
in Eq.~(\ref{4-1}) is used in the drift kinetic 
theory~\cite{Hinton1976,H&S,Helander} where 
the neoclassical transport fluxes are calculated from 
the first-order distribution function given as the solution 
of the drift kinetic equation [see Eq.~(\ref{C1-5})]. 
In the gyrokinetic theory, small amplitudes of fluctuations of 
${\cal O}(\epsilon)$ are assumed so that the fluctuation parts 
appear from the first order as seen below.  
%}
%===

\subsection{Zeroth-order flows}

The zeroth-order part $f_{a0}$ of the 
distribution function $f_a$ in $\epsilon$ is considered to 
represent the equilibrium part which 
contains no fluctuations, and we accordingly write 
\begin{equation}
\label{4A-1}
f_{a0} =  \langle f_{a0} \rangle_{\rm ens} , 
\hspace*{5mm} 
\widehat{f}_{a0} =  0
.
\end{equation}
The zeroth-order density $n_{a0}^{(g)}$ is given by
\begin{equation}
\label{4A-2}
n_{a0}^{(g)} ({\bf x}, t)  \equiv \int d^6 Z
\; D_{a0} ({\bf X}, t) f_{a0} ({\bf Z}, t) 
\delta^3 ( {\bf X} - {\bf x} )
,
\end{equation}
where $D_{a0}$ represents the zeroth-order Jacobian given by 
\begin{equation}
\label{4A-3}
D_{a0}({\bf X}, t)
= 
\frac{B({\bf X}, t)}{m_a}
.
\end{equation}
The zeroth-order part 
$({d {\bf X}}/dt )_0$ of the gyrocenter velocity 
${d {\bf X}}/dt$ 
is given by 
Eq.~(\ref{B-1}) and  
%
%\begin{equation}
%\label{4A-4}
%\left(
%\frac{d {\bf X}}{dt} 
%\right)_0
%= 
%U{\bf b} ({\bf X}, t) 
%, 
%\end{equation}
%
it has only the component parallel to the background magnetic field. 

Noting that $f_{a0}$ is independent of the gyrophase angle $\xi$ and 
using Eqs.~(\ref{3-4}), (\ref{3-7}), (\ref{B-1}), and (\ref{B-6})
we have  
\begin{eqnarray}
\label{4A-5}
\left(
\frac{1}{e_a}\frac{\partial {\bf P}_a({\bf x}, t)}{\partial t}
\right)_0
& = & 
\int d^6 Z  \; 
D_{a0} f_{a0} 
\delta^3 ({\bf X} - {\bf x} )
\left(
\frac{d \boldsymbol{\rho}_a}{dt} 
\right)_0
\nonumber \\ 
& = & 
0
,
\end{eqnarray}
and 
\begin{equation}
\label{4A-6}
\left(
\frac{c}{e_a} {\bf M}_a({\bf x}, t)
\right)_0
= 0
.
\end{equation}
Thus, the polarization and magnetization never produce 
particle flows of ${\cal O}(n_{a0} v_{Ta})$. 
From Eqs.~(\ref{2-15}) and (\ref{2-15*}), we also have
\begin{equation}
\label{4A-7}
\boldsymbol{\Gamma}^C_{a0} ({\bf x}, t)
= \boldsymbol{\Gamma}^{C*}_{a0} ({\bf x}, t)
= 0
.
\end{equation}

In the present work, 
we use the low-flow ordering in which 
the lowest-order flow velocity is in the order of 
${\cal O}(\epsilon v_{Ta})$. 
This means that the zeroth-order particle flow vanishes,
\begin{equation}
\label{4A-8}
\boldsymbol{\Gamma}^{(p)}_{a0} ({\bf x}, t)
= 0
, 
\end{equation}
and the 
zeroth-order gyrocenter flow given by $f_{a0}$ also 
vanishes,
\begin{eqnarray}
\label{4A-9}
& & \boldsymbol{\Gamma}^{(g)}_{a0} ({\bf x}, t)  
  \equiv  
n^{(g)}_{a0} {\bf u}^{(g)}_{a0} ({\bf x}, t) 
\nonumber \\ 
& \equiv  & 
\int d^6 Z
\; D_{a0} ({\bf Z}, t) f_{a0} ({\bf Z}, t) 
\delta^3 ( {\bf X} - {\bf x} )
\left(
\frac{d {\bf X}}{dt} 
\right)_0
= 0
.
\hspace*{5mm}
\end{eqnarray}

\subsection{First-order flows}

In the first-order in $\epsilon$, 
the gyrocenter distribution function generally consists of 
ensemble average and fluctuation parts, 
\begin{equation}
\label{4B-1}
f_{a1} =  \langle f_{a1} \rangle_{\rm ens} + \widehat{f}_{a1}
.
\end{equation}
In the same way, the first-order particle and gyrocenter 
flows are written as
\begin{equation}
\label{4B-2}
\boldsymbol{\Gamma}^{(p)}_{a1} ({\bf x}, t)
= 
\langle
\boldsymbol{\Gamma}^{(p)}_{a1} ({\bf x}, t)
\rangle_{\rm ens}
+ \widehat{\boldsymbol{\Gamma}}^{(p)}_{a1} ({\bf x}, t)
,
\end{equation}
and
\begin{equation}
\label{4B-3}
\boldsymbol{\Gamma}^{(g)}_{a1} ({\bf x}, t)
= 
\langle
\boldsymbol{\Gamma}^{(g)}_{a1} ({\bf x}, t)
\rangle_{\rm ens}
+ \widehat{\boldsymbol{\Gamma}}^{(g)}_{a1} ({\bf x}, t)
,
\end{equation}
respectively. 
As explained in Appendix~C, 
the collision term vanishes to the zeroth order in $\epsilon$ and 
it is regarded as of the first order. 
Then we see from Eqs.~(\ref{2-15}) and (\ref{2-15*}) that 
$\boldsymbol{\Gamma}_a^C$ and   
$\boldsymbol{\Gamma}_a^{C*}$ are of ${\cal O}(\epsilon^2 n_{a0} v_{Ta})$ 
($n_{a0}$: the background particle density)
and 
\begin{equation}
\label{4B-4}
\boldsymbol{\Gamma}_{a1}^C 
= 
\boldsymbol{\Gamma}_{a1}^{C*}
= 
0
.
\end{equation}

\subsubsection{Ensemble-averaged part}

The first-order ensemble-averaged gyrocenter flow is 
written as 
\begin{eqnarray}
\label{4B1-1}
\langle
\boldsymbol{\Gamma}^{(g)}_{a1} ({\bf x}, t)
\rangle_{\rm ens}
& = & 
\int d^6 Z
\; \delta^3 ( {\bf X} - {\bf x} )
\left[
D_{a0} f_{a0} 
\left\langle
\left(
\frac{d {\bf X}}{dt} 
\right)_1
\right\rangle_{\rm ens}
\right.
\nonumber \\ & & 
\left. \mbox{}
+ ( D_{a0} \langle f_{a1} \rangle_{\rm ens} + D_{a1} f_{a0} ) 
\left(
\frac{d {\bf X}}{dt} 
\right)_0
\right]
,
\end{eqnarray}
where $(d {\bf X}/dt)_0$ and 
$\langle (d {\bf X}/dt)_1\rangle_{\rm ens}$ are given as 
functions of $({\bf Z}, t)$ as shown in by 
Eqs.~(\ref{B-1}) and (\ref{B-3}), 
respectively. 
It is found from Eq.~(\ref{3-2}) that the first-order polarization flow vanishes,
\begin{equation}
\label{4B1-2}
\langle
\boldsymbol{\Gamma}^{\rm pol}_{a1} ({\bf x}, t)
\rangle_{\rm ens}
\equiv 
\left\langle
\left(
\frac{1}{e_a}\frac{\partial {\bf P}_a ({\bf x}, t)}{\partial t}
\right)_1
\right\rangle_{\rm ens}
= 0
.
\end{equation}
From Eq.~(\ref{3-7}), we obtain  
\begin{equation}
\label{4B1-3}
\left\langle
\left(
\frac{c}{e_a}  {\bf M}_a({\bf x}, t)
\right)_1
\right\rangle_{\rm ens}
= 
- \frac{c}{e_a B} (P_{a\perp})_0 {\bf b}
,
\end{equation}
and the first-order magnetization flow, 
\begin{eqnarray}
\label{4B1-4}
\langle
\boldsymbol{\Gamma}^{\rm mag}_{a1} ({\bf x}, t)
\rangle_{\rm ens}
& \equiv & 
\nabla \times
\left\langle
\left(
\frac{c}{e_a}  {\bf M}_a
\right)_1
\right\rangle_{\rm ens}
\nonumber \\
& = & 
- \nabla \times
\left(
\frac{c}{e_a B} (P_{a\perp})_0 {\bf b}
\right)
, 
\end{eqnarray}
where 
\begin{equation}
\label{4B1-5}
(P_{a\perp})_0
\equiv 
\int d^6 Z
\; \delta^3 ( {\bf X} - {\bf x} )
D_{a0} f_{a0} \, \mu B
.
\end{equation}

Using Eqs.~(\ref{3-6}), (\ref{4B-4}), (\ref{4B1-1}), 
(\ref{4B1-2}), and (\ref{4B1-4}), 
the total first-order ensemble-averaged particle flow is written as 
\begin{eqnarray}
\label{4B1-6}
& & 
\langle
\boldsymbol{\Gamma}^{(p)}_{a1} ({\bf x}, t)
\rangle_{\rm ens}
=
\langle
\boldsymbol{\Gamma}^{(g)}_{a1} 
\rangle_{\rm ens}
+ \langle
\boldsymbol{\Gamma}^{\rm mag}_{a1} 
\rangle_{\rm ens}
\nonumber \\ 
&   & =
\int d^6 Z
\; \delta^3 ( {\bf X} - {\bf x} ) 
D_{a0} \langle f_{a1} \rangle_{\rm ens}
 U {\bf b} 
+ \frac{n_{a0}}{B}  
\langle {\bf E}_1 \rangle_{\rm ens}
\times {\bf b}
\nonumber \\ & & 
\mbox{} 
\hspace*{5mm}
+  \frac{c}{e_a B} 
\left[ {\bf b} \times
\nabla (P_{a\perp})_0 
+ \bigl\{ 
(P_{a\parallel})_0  - (P_{a\perp})_0 
\bigl\}
\left( \nabla \times {\bf b} \right)
\right]
\nonumber \\ 
&  & =
{\bf b} 
\biggl[
\int d^6 Z
\; \delta^3 ( {\bf X} - {\bf x} ) 
D_{a0} \langle f_{a1} \rangle_{\rm ens}
 U
 + \frac{c}{e_a B} 
\bigl\{ (P_{a\parallel})_0 
\nonumber \\ & & 
\mbox{} 
\hspace*{5mm}
 - (P_{a\perp})_0 \bigr\}
\left( {\bf b} \cdot \nabla \times {\bf b} \right)
\biggr]
+  \frac{c}{e_a B} 
\Bigl[
n_{a0} e_a \langle {\bf E}_1 \rangle_{\rm ens}
\nonumber \\ & & 
\mbox{} 
\hspace*{5mm}
-
\nabla \cdot 
\left\{ (P_{a\parallel})_0 {\bf b} {\bf b} 
\mbox{} 
+ (P_{a\perp})_0 
( {\bf I} - {\bf b} {\bf b} )
\right\}
\Bigr]
 \times {\bf b} 
.
\end{eqnarray}
where 
\begin{equation}
\label{4B1-7}
(P_{a\parallel})_0
\equiv 
\int d^6 Z
\; \delta^3 ( {\bf X} - {\bf x} )
D_{a0} f_{a0} \frac{1}{2} m_a U^2
.
\end{equation}
In a case where, as described in Appendix~C.1, 
 $f_{a0}$ takes the form of the local Maxwellian distribution with 
no mean flow, 
the zeroth-order pressure is isotropic so that we can 
write 
$(P_{a\parallel})_0 = (P_{a\perp})_0 = P_{a0}$. 
Equation~(\ref{4B1-6}) agrees with the 
magnetization law in 
drift kinetics.~\cite{magnetization_law}

Within accuracy up to 
${\cal O}(\epsilon n_{a0} v_{Ta})$,  
Eq.~(\ref{4B1-6}) is rewritten more compactly as 
\begin{eqnarray}
\label{4B1-8}
\langle
\boldsymbol{\Gamma}^{(p)}_{a1} ({\bf x}, t)
\rangle_{\rm ens}
& = & 
\int d^6 Z
\; \delta^3 ({\bf X} + 
\boldsymbol{\rho}_{a1} - {\bf x} )
\nonumber \\
& & 
\mbox{}
\times
\left[
D_a \langle f_a \rangle_{\rm ens} {\bf v}_c 
+
D_{a0} 
 f_{a0} 
{\bf v}_{da}
\right]
,   
\hspace*{5mm}
\end{eqnarray}
where ${\bf v}_c$ and ${\bf v}_{da}$ are given by 
Eqs.~(\ref{A-13}) and (\ref{B-3}), respectively, 
and $\boldsymbol{\rho}_{a1}$ represents the lowest-order 
(or first-order) expression of the gyroradius vector shown in 
Eq.~(\ref{A-12}).   
In the first term of the integrand on the right-hand side of 
Eq.~(\ref{4B1-8}), 
we need to use 
$D_a = D_{a0} + D_{a1}$ and $f_a = f_{a0} + f_{a1}$ 
in order to keep the validity up to ${\cal O}(\epsilon n_{a0} v_{Ta})$.

\subsubsection{Turbulent part}

The first-order turbulent gyrocenter flow is given from 
Eq.~(\ref{2-14}) as 
\begin{equation}
\label{4B2-1}
\widehat{\boldsymbol{\Gamma}}^{(g)}_{a1} ({\bf x}, t)
=
\int d^6 Z
\; D_0 \delta^3 ( {\bf X} - {\bf x} )
(
\widehat{f}_{a1}
U {\bf b}
+ f_{a0}
\widehat{\bf v}_{ga}
)
,
\end{equation}
where the first-order turbulent gyrocenter velocity $\widehat{\bf v}_{ga}$ 
is given by Eq.~(\ref{B-4}). 
The first-order turbulent polarization flow is derived from 
Eq.~(\ref{3-4}) as
\begin{eqnarray}
\label{4B2-2}
& & \widehat{\boldsymbol{\Gamma}}^{\rm pol}_{a1} ({\bf x}, t)
\nonumber \\ & & 
\mbox{}
=
\sum_{l=0}^\infty 
\frac{(-1)^{l+1}}{(l+1)!} \frac{\partial^l}{\partial x^{j_1}\cdots 
\partial x^{j_l}}
\biggl[
\int d^6 Z
\;  \delta^3 ({\bf X} - {\bf x} ) D_{a0}
\nonumber \\ & & 
\mbox{}
\hspace*{5mm} 
\cdot
\biggl\{
\rho_{a1}^{j_1} 
\cdots \rho_{a1}^{j_l}
f_{a0} 
\biggl(
 \frac{e_a}{m_a c} 
 \widehat{\bf A}_\perp 
+ \frac{c}{B} {\bf b} 
\times \nabla 
%===revision===
%\textcolor{red}{
\widehat{\psi}_a
%}
%===
\biggr) 
\nonumber \\ & & 
\mbox{}
\hspace*{5mm} 
- \rho_{a1}^{j_1} 
 \cdots 
\rho_{a1}^{j_{l-1}}
\frac{e_a}{B} 
%===revision===
%\textcolor{red}{
\widetilde{\widehat{\psi}}_a
%}
%===
\frac{\partial f_{a0}}{\partial \mu}
\left(
 \rho_{a1}^{j_l} {\bf v}_{c\perp}
+ l (v_{c\perp})^{j_l}
\boldsymbol{\rho}_{a1} 
\right)
\biggl\}
\biggr],
\nonumber \\ & & 
\end{eqnarray}
where $\rho_{a1}^{j}$ is the $j$th component of 
$\boldsymbol{\rho}_{a1}$. 
On the right-hand side of Eq.~(\ref{4B2-2}), 
%===revision===
%\textcolor{red}{
$\widetilde{\widehat{\psi}}_a
\equiv \widehat{\psi}_a - 
\langle \widehat{\psi}_a \rangle_\xi$ 
%}
%===
is the gyrophase-dependent part 
of 
%===revision===
%\textcolor{red}{
$\widehat{\psi}_a \equiv \psi_a - \langle \psi_a \rangle_{\rm ens} 
\equiv \widehat{\phi} - c^{-1} {\bf v}_c \cdot \widehat{\bf A}$
%}
%===
where $\widehat{\phi}$ and $\widehat{\bf A}$ should be 
evaluated at ${\bf X} + \boldsymbol{\rho}_{a1}$. 
The first-order turbulent magnetization flow is derived from 
Eq.~(\ref{3-7}) as
\begin{eqnarray}
\label{4B2-3}
& & 
\widehat{\boldsymbol{\Gamma}}^{\rm mag}_{a1}  ({\bf x}, t)
= 
\left(
\frac{c}{e_a} \nabla \times {\bf M}_a
\right)_1
\nonumber \\ & & 
= 
\sum_{l=1}^\infty 
\frac{(-1)^l}{l!} \frac{\partial^l}{\partial x^{j_1}\cdots 
\partial x^{j_l}}
\biggl(
\int d^6 Z
\;  \delta^3 ({\bf X} - {\bf x} )
D_{a0} 
\nonumber \\ & & 
\mbox{}
\hspace*{5mm} 
\cdot 
\biggl[
\; \rho_{a1}^{j_1} 
 \cdots 
\rho_{a1}^{j_l}
\biggl\{ \widehat{f}_{a1} \; {\bf v}_c 
- f_{a0} \frac{e_a}{m_a c} \biggl( \widehat{A}_\parallel {\bf b}
+ \frac{l}{(l+1)}  \widehat{\bf A}_\perp \biggr)
\nonumber \\ & & 
\mbox{}
\hspace*{5mm} 
+ \frac{1}{(l+1)} f_{a0} \frac{c}{B} {\bf b} 
\times \nabla 
%===revision===
%\textcolor{red}{
\widehat{\psi}_a
%}
%===
\biggr\}
+ \rho_{a1}^{j_1} 
 \cdots 
\rho_{a1}^{j_{l-1}}
\frac{e_a}{B} \widetilde{\widehat{\psi}}_a
\frac{\partial f_{a0}}{\partial \mu}
\nonumber \\ & & 
\mbox{}
\hspace*{5mm}
\cdot
\biggl\{
\rho_{a1}^{j_l}
U {\bf b} 
+ \frac{l}{l+1} 
\biggl(
\rho_{a1}^{j_l}
{\bf v}_{c\perp}
-
(v_{c\perp})^{j_l}
\boldsymbol{\rho}_{a1}
\biggr)
\biggr\}
\biggr]
\biggr)
.
\end{eqnarray}
Then, using Eqs.~(\ref{4B2-1})--(\ref{4B2-3}), 
the first-order turbulent particle flux is written as 
\begin{eqnarray}
\label{4B2-4}
& &  \widehat{\boldsymbol{\Gamma}}^{(p)}_{a1} ({\bf x}, t) 
\equiv 
\widehat{\boldsymbol{\Gamma}}^{(g)}_{a1} 
+ \widehat{\boldsymbol{\Gamma}}^{\rm pol}_{a1}
+ \widehat{\boldsymbol{\Gamma}}^{\rm mag}_{a1}
\nonumber \\
& & 
\mbox{}
= 
\sum_{l=0}^\infty 
\frac{(-1)^l}{l!} \frac{\partial^l}{\partial x^{j_1}\cdots 
\partial x^{j_l}}
\biggl[
\int d^6 Z
\;  \delta^3 ({\bf X} - {\bf x} )
D_{a0} 
\nonumber \\
& & 
\mbox{}
\hspace*{5mm}
\cdot
\rho_{a1}^{j_1} 
 \cdots 
\rho_{a1}^{j_l}
\Biggl\{
 \widehat{f}_{a1}  {\bf v}_c 
+
\left( 
- f_{a0} 
\frac{e_a}{m_a c} \widehat{\bf A}
+
\frac{e_a \widetilde{\widehat{\psi}}_a}{B} 
\frac{\partial f_{a0}}{\partial \mu} {\bf v}_c 
\right)
\Biggr\}
\Biggr]
\nonumber \\
& & 
\mbox{}
= 
\int d^6 Z
\; \delta^3 ({\bf X} + 
\boldsymbol{\rho}_{a1} - {\bf x} ) 
D_{a0} 
\nonumber \\
& & 
\mbox{}
\hspace*{5mm}
\cdot
\Biggl[
 \widehat{f}_{a1}  {\bf v}_c 
+
\biggl(
- f_{a0} 
\frac{e_a}{m_a c} \widehat{\bf A}
+
\frac{e_a \widetilde{\widehat{\psi}}_a}{B} 
\frac{\partial f_{a0}}{\partial \mu} {\bf v}_c 
\biggr)
\Biggr]
+ {\cal O}(\epsilon^2 n_{a0} v_{Ta})
. 
\nonumber \\
& & 
\end{eqnarray}

Summing up Eqs.~(\ref{4B1-8}) and (\ref{4B2-4}), 
we obtain the expression of the first-order 
particle flow, which is valid up to ${\cal O}(\epsilon n_{a0} v_{Ta})$, as 
\begin{eqnarray}
\label{4B2-5}
& &
\hspace*{-3mm}
 \boldsymbol{\Gamma}_a^{(p)} ({\bf x}, t)
= \langle
\boldsymbol{\Gamma}_a^{(p)} ({\bf x}, t)
\rangle_{\rm ens}
+ \widehat{\boldsymbol{\Gamma}}_a^{(p)} ({\bf x}, t)
\nonumber \\ & & \mbox{}
\hspace*{-3mm}
=
\int d^6 Z
\; \delta^3 ({\bf X} + 
\boldsymbol{\rho}_{a1} - {\bf x} )
\Biggl[
D_a ({\bf Z}, t) f_a ({\bf Z}, t) {\bf v}_c 
\nonumber \\ & &
\mbox{}
+
D_{a0} 
\left\{ 
 f_{a0} 
\left( 
%===revision===
%\textcolor{red}{
{\bf v}_{Ba}
%}
%===
- \frac{e_a}{m_a c} \widehat{\bf A}
\right)
+
\frac{e_a \widetilde{\psi}_a}{B} 
\frac{\partial f_{a0}}{\partial \mu} {\bf v}_c 
\right\}
\Biggr]
.
\nonumber \\ & &
\mbox{}
\end{eqnarray}
where 
%===revision===
%\textcolor{red}{
${\bf v}_{Ba}$
%}
%===
is defined by Eq.~(\ref{A-26}). 
In the same way as in Eq.~(\ref{4B1-8}), 
$D_a = D_{a0} + D_{a1}$ and $f_a = f_{a0} + f_{a1}$ 
should be used in the first term of the integrand 
on the right-hand side of Eq.~(\ref{4B2-5}), 
in order to keep the validity up to ${\cal O}(\epsilon n_{a0} v_{Ta})$.

\subsection{Second-order flows}

When considering particle confinement of magnetically 
confined plasmas on the transport time scale 
of $(\epsilon^2 \omega_{Ta})^{-1}$, 
it is important to evaluate the ensemble-averaged or 
mean particle flux across the surface formed by field lines. 
We find from Eq.~(\ref{3-2}) that the second-order 
Ensemble-averaged polarization flow vanishes,
\begin{equation}
\label{4C-1}
\left\langle
\boldsymbol{\Gamma}^{\rm pol}_{a2} ({\bf x}, t) 
\right\rangle_{\rm ens}
\equiv
\left\langle
\left(
\frac{1}{e_a}\frac{\partial {\bf P}_a}{\partial t}
\right)_2
\right\rangle_{\rm ens}
= 0
,
\end{equation}
as well as the zeroth- and first-order parts shown 
in Eqs.~(\ref{4A-5}) and (\ref{4B1-2}).
 
For plasmas confined in the toroidal magnetic configuration 
where the zeroth-order equilibrium distribution function 
$F_{a0}$ is given by the Maxwellian with no mean flow, 
we see from Eq.~(\ref{3-7}) that the second-order ensemble-averaged 
magnetization flow is given by
\begin{eqnarray}
\label{4C-2}
\left\langle
\boldsymbol{\Gamma}^{\rm mag}_{a2} ({\bf x}, t) 
\right\rangle_{\rm ens}
& \equiv & 
\left\langle
\left(
\frac{c}{e_a} \nabla \times {\bf M}_a
\right)_2
\right\rangle_{\rm ens}
\nonumber \\
& = &
- \nabla \times
\left(
\frac{c}{e_a B} (P_\perp)_{a1} {\bf b}
\right)
,
\end{eqnarray}
where 
$
(P_\perp)_{a1} \equiv 
\int d^6 Z
\; \delta^3 ( {\bf X} - {\bf x} ) 
D_{a0} \langle f_{a1} \rangle_{\rm ens} \mu B
$.
For this Maxwellian equilibrium distribution function $f_{a0}$, 
we have the scalar equilibrium pressure 
$P_{a0} = (P_{a\parallel})_0 = (P_{a\perp})_0 $ and 
the average electrostatic potential 
$\langle \phi \rangle_{\rm ens}$ which are given as 
flux surface functions, as explained after Eq.~(\ref{C1-4}) 
in Appendix~C. 
Then the first-order ensemble-averaged particle flow in 
Eq.~(\ref{4B1-8}) 
is rewritten as 
\begin{eqnarray}
\label{4C-3}
\langle
\boldsymbol{\Gamma}^{(p)}_{a1} ({\bf x}, t)
\rangle_{\rm ens}
 & = &
\int d^6 Z
\; \delta^3 ( {\bf X} - {\bf x} ) 
D_{a0} \langle f_{a1} \rangle_{\rm ens}
 U {\bf b} 
\nonumber \\ & & 
\mbox{} 
+  \frac{c}{e_a B} 
\bigl(
n_{a0} e_a \langle {\bf E}_1 \rangle_{\rm ens}
-
\nabla  
P_{a0}
\bigr)
 \times {\bf b} 
,
\hspace*{5mm}
\end{eqnarray}
which has no component in the radial direction perpendicular to 
the magnetic flux surface, because 
$\langle {\bf E}_1 \rangle_{\rm ens} 
= - \nabla \langle \phi \rangle_{\rm ens}$
and $\nabla P_{a0}$ are both perpendicular to the surface. 
Therefore the mean radial particle flow is of 
${\cal O}(\epsilon^2 n_{a0} v_{Ta})$, 
which is consistent with 
the ordering of the transport time scale given by $(\epsilon^2 \omega_{Ta})^{-1}$.

The second-order ensemble-averaged gyrocenter flow is obtained from 
Eq.~(\ref{2-14}) 
as
\begin{eqnarray}
\label{4C-4}
\langle
\boldsymbol{\Gamma}^{(g)}_{a2} ({\bf x}, t)
\rangle_{\rm ens}
 & = &
\int d^6 Z
\; \delta^3 ( {\bf X} - {\bf x} )
D_{a0} 
\biggl[
 f_{a0} 
{\bf v}_{da2}
\nonumber \\ & & 
\mbox{} 
+ 
 \langle f_{a1} \rangle_{\rm ens}
{\bf v}_{da}
+ 
\langle
\,
\widehat{f}_{a1} 
\,
\widehat{\bf v}_{ga}
\rangle_{\rm ens}
\nonumber \\ & & 
\mbox{}
+ \left\langle 
f_{a2} + \frac{D_{a1}}{D_{a0}} f_{a1} 
\right\rangle_{\rm ens}
 U {\bf b}
\biggr]
,
\hspace*{5mm}
\end{eqnarray}
where ${\bf v}_{da}$,  
$\widehat{\bf v}_{ga}$, and ${\bf v}_{da2}$ 
are given by 
Eqs.~(\ref{B-3}), (\ref{B-4}), and (\ref{B-5}), 
respectively.

The remaining part of the second-order ensemble-averaged particle flow 
is derived using Eq.~(\ref{2-15}) and (\ref{2-15*}) as 
\begin{eqnarray}
\label{4C-5}
& & 
\langle
\boldsymbol{\Gamma}^C_{a2} ({\bf x}, t)
\rangle_{\rm ens}
=
\langle
\boldsymbol{\Gamma}^{C*}_{a2} ({\bf x}, t)
\rangle_{\rm ens}
\nonumber \\
& & =
\int d^6 z' 
\; D_a^{(p)}  \langle (\widetilde{C}_a^{(p)})_1  \rangle_{\rm ens}
\delta^3 ( {\bf x}' - {\bf x} ) 
\frac{{\bf v}' \times {\bf b}}{\Omega_a}
\nonumber \\
& & =
\frac{c}{e_a B} {\bf F}_{a1} \times {\bf b}
\nonumber \\ 
& & =
\int d^6 z'
\; D_a^{(p)} 
\; \langle \widetilde{f}_{a2} \rangle_{\rm ens}
\;
\delta^3 ( {\bf x}' - {\bf x} )
\;
{\bf v}'_\perp
, 
\end{eqnarray}
where  
$\langle (\widetilde{C}_a^{(p)})_1  \rangle_{\rm ens}$ 
is defined by
\begin{equation}
\label{4C-6}
\langle (\widetilde{C}_a^{(p)})_1  \rangle_{\rm ens}
\equiv \sum_b
\bigl\{
C_{ab}^{(p)}
[\langle \widetilde{f}_{a1}^{(p)} \rangle_{\rm ens}, f_{b0}]
+ C_{ab}^{(p)}
[f_{a0}, \langle \widetilde{f}_{b1}^{(p)} \rangle_{\rm ens}]
\bigr\}
,
\end{equation}
${\bf F}_{a1}$ is the collisional friction force defined by 
\begin{equation}
\label{4C-7}
{\bf F}_{a1} 
\equiv
\int d^6 z' \; D_a^{(p)}  \delta^3 ({\bf x}' - {\bf x})
\langle (C_a^{(p)})_1 \rangle_{\rm ens}
\;
m_a {\bf v}'
,
\end{equation}
and $\widetilde{f}_{a2}$ is obtained using Eq.~(\ref{2-7}). 
It is verified from Eq.~(\ref{4C-5}) that 
$\langle \boldsymbol{\Gamma}^C_{a2} \rangle_{\rm ens}
= \langle \boldsymbol{\Gamma}^{C*}_{a2} \rangle_{\rm ens}$ 
represents the classical collisional 
particle flow.~\cite{Hinton1976,H&S,Helander}

As seen from Eqs.~(\ref{2-18}) and (\ref{4C-1}), 
the total second-order particle flow is given by the sum of 
the gyrocenter, magnetization, and classical particle flows, 
\begin{equation}
\label{4C-8}
\langle
\boldsymbol{\Gamma}^{(p)}_{a2} ({\bf x}, t)
\rangle_{\rm ens}
=
\langle
\boldsymbol{\Gamma}^{(g)}_{a2} 
\rangle_{\rm ens}
+
\langle
\boldsymbol{\Gamma}^{\rm mag}_{a2} 
\rangle_{\rm ens}
+
\langle
\boldsymbol{\Gamma}^C_{a2} 
\rangle_{\rm ens}
.
\end{equation}
It is recalled here that 
the tangential component of the mean particle flow to the magnetic flux surface 
is dominated by the first-order flow 
$\langle \boldsymbol{\Gamma}^{(p)}_{a1} \rangle_{\rm ens}$ given in 
Eq.~(\ref{4B1-6}) 
although the normal component is of the second order. 
Now using Eqs.~(\ref{3-7}), (\ref{4C-4}), 
(\ref{4C-5}), and (\ref{4C-8}), 
the component of the second-order particle flow 
$\langle
\boldsymbol{\Gamma}^{(p)}_{a2}
\rangle_{\rm ens}$
perpendicular to the background magnetic field line 
is given by  
\begin{eqnarray}
\label{4C-9}
& & 
\langle
\boldsymbol{\Gamma}^{(p)}_{a \perp 2} ({\bf x}, t)
\rangle_{\rm ens}
=
\langle
\boldsymbol{\Gamma}^{(g)}_{a \perp a2}
\rangle_{\rm ens}
+
\langle
\boldsymbol{\Gamma}^{\rm mag}_{a \perp 2}
\rangle_{\rm ens}
+
\langle
\boldsymbol{\Gamma}^C_{a2} 
\rangle_{\rm ens}
\nonumber \\ 
&  & 
=
\int d^6 Z
\; \delta^3 ( {\bf X} - {\bf x} )
D_{a0} 
\left[
 \langle f_{a1} \rangle_{\rm ens}
{\bf v}_{da}
+ 
\langle
\,
\widehat{f}_{a1}
\,
(\widehat{\bf v}_{ga})_\perp
\rangle_{\rm ens}
\right]
\nonumber \\ 
&  & 
\mbox{}
\hspace*{2mm}
- 
\left[ \nabla \times
\left(
\frac{c}{eB} (P_{a\perp})_1 {\bf b}
\right)
\right]_\perp
+  \frac{c}{e_a B} 
\left[ 
n_{a0} e
\langle {\bf E}_2 \rangle_{\rm ens}
+
{\bf F}_{a1} 
\right] \times {\bf b}
\nonumber \\ &  & =
\mbox{} 
\frac{c}{e_a B} 
\Bigl[
-
\nabla \cdot 
\left\{ (P_{a\parallel})_1 {\bf b} {\bf b} 
+ (P_{a\perp})_1 
( {\bf I} - {\bf b} {\bf b} )
\right\}
\nonumber \\ &   & 
\mbox{} 
\hspace*{8mm}
+ n_{a1} e_a \langle {\bf E}_1 \rangle_{\rm ens}
+
n_{a0} e
\langle {\bf E}_2 \rangle_{\rm ens}
+
{\bf F}_{a1} 
\nonumber \\ &   & 
\mbox{} 
\hspace*{8mm}
- \frac{c}{B}
\int d^6 Z
\; \delta^3 ( {\bf X} - {\bf x} )
D_{a0} 
\langle
\,
\widehat{f}_{a1}
\,
\nabla \langle \widehat{\psi}_a \rangle_\xi
\rangle_{\rm ens}
\Bigr]
 \times {\bf b} 
,
\hspace*{2mm}
\end{eqnarray}
where 
$
n_{a1} \equiv 
\int d^6 Z
\; \delta^3 ( {\bf X} - {\bf x} ) 
D_{a0} \langle f_{a1} \rangle_{\rm ens} 
$,
$
(P_{a\parallel})_1 \equiv 
\int d^6 Z
\; \delta^3 ( {\bf X} - {\bf x} ) 
D_{a0} \langle f_{a1} \rangle_{\rm ens} m_a U^2
$,
$
\langle {\bf E}_1 \rangle_{\rm ens}
= -\nabla \langle \phi_1 \rangle_{\rm ens}
$,
and
$
\langle {\bf E}_2 \rangle_{\rm ens}
= -\nabla \langle \phi_2 \rangle_{\rm ens}
- c^{-1} \partial {\bf A}/\partial t
$
are used. 
In toroidal confinement systems,  
the lowest-order ensemble-averaged electrostatic potential 
$\langle \phi_1 \rangle_{\rm ens}$ is considered to be 
uniform over the magnetic flux surface. 
On the right-hand side of Eq.~(\ref{4C-9}),
the part including the anisotropic pressure tensor   
represents the neoclassical particle 
transport~\cite{Hinton1976,H&S,Helander}
while the turbulent particle transport is given by 
the last term including the correlation between 
the fluctuating distribution function 
and the gradient of the gyrophase-averaged 
fluctuating potential field.~\cite{Sugama1996}

\section{LAGRANGIAN FOR VARIATIONAL DERIVATION OF POISSON'S EQUATION AND AMP\`{E}RE'S LAW}

The action integral for the gyrokinetic Vlasov-Poisson-Amp\`{e}re system is given by 
\begin{equation}
\label{5-1}
I  \equiv  \int_{t_1}^{t_2} dt \; L_{GKF} 
\equiv  \int_{t_1}^{t_2} dt \; ( L_{GK} + L_F ) 
,
\end{equation}
where the Lagrangian $L_{GK}$ is written as
\begin{equation}
\label{5-2}
L_{GK}
\equiv 
L_{GK0} + L_{GK1} + L_{GK2} 
.
\end{equation}
Here, we use the gyrocenter distribution function $f_a$ to define 
$L_{GK0}$ and $L_{GK1}$ by
\begin{eqnarray}
\label{5-3}
& & \left[
\begin{array}{c}
L_{GK0}  \\
L_{GK1} 
\end{array}
\right]
 \equiv 
\sum_a
\int d^6 Z_0
\; 
D_a ({\bf Z}_0, t_0) f_a ({\bf Z}_0, t_0) 
\nonumber \\
& & 
\hspace*{30mm}
\mbox{} \times
\left[
\begin{array}{c}
L_{GYa 0}  ({\bf Z}_a (t), \dot{\bf Z}_a (t), t)  \\
L_{GYa 1}  ({\bf Z}_a (t), t)
\end{array}
\right]
\nonumber \\
& &
\equiv 
\sum_a
\int d^6 Z
\; 
D_a ({\bf Z}, t) f_a ({\bf Z}, t) 
\left[
\begin{array}{c}
L_{GYa 0}  ({\bf Z}, \dot{\bf Z}, t)  \\
L_{GYa 1}  ({\bf Z}, t)
\end{array}
\right]
,
\end{eqnarray}
where the gyrocenter 
phase-space orbit for the particle species $a$ is 
represented by ${\bf Z}_a (t)$ which satisfies 
the initial condition ${\bf Z}_a (t_0) = {\bf Z}_0$. 
The gyrocenter Lagrangian $L_{GYa 0}$ appearing in Eq.~(\ref{5-3}) is defined by
\begin{eqnarray}
\label{5-4}
L_{GYa 0}  ({\bf Z}, \dot{\bf Z}, t) 
 & \equiv & 
\frac{e_a}{c} {\bf A}^*_a ({\bf X}, U, t)
\cdot \dot{\bf X}
+ \frac{m_a c}{e_a} \mu \; \dot{\xi}
\nonumber \\ 
& & 
\mbox{}
- 
\biggl(
\frac{1}{2} m_a U^2 + \mu B ({\bf X}, t)
\biggr)
,
\end{eqnarray}
which describes the gyrocenter motion for the case 
where the electrostatic potential $\phi$ and
the vector potential fluctuation $\widehat{\bf A}$ 
vanish. 
In this section, we use the modified vector potential 
$
{\bf A}_a^* ({\bf X}, U, t)
 \equiv  
{\bf A} ({\bf X}, t)
+ (m_a c/e_a) U {\bf b} ({\bf X}, t)
$
which is obtained from Eq.~(\ref{A-21}) with 
the second-order small term neglected. 
%
%\begin{equation}
%H_{GYa 0}  ({\bf Z}, t)  \equiv  
%\frac{1}{2} m_a U^2 + \mu B ({\bf X}, t)
%\end{equation}
%
The gyrocenter Lagrangian $L_{GYa 1}$ is the part which 
linearly depends on $\phi$ and $\widehat{\bf A}$,
\begin{eqnarray}
\label{5-5}
& & L_{GYa 1}  ({\bf Z}, t) 
% \equiv  - H_{GYa 1}  ({\bf Z}, t) 
 \equiv  
- e_a \langle
\psi_a ({\bf Z}, t) 
\rangle_\xi
\nonumber \\
& & 
\equiv
- e_a \left\langle
\phi ({\bf X} + \boldsymbol{\rho}_{a1}, t) 
- \frac{{\bf v}_c}{c} \cdot 
\widehat{\bf A}
({\bf X} + \boldsymbol{\rho}_{a1}, t) 
\right\rangle_\xi
.
\end{eqnarray}
The second-order Lagrangian $L_{GK2}$ is given by 
\begin{equation}
\label{5-6}
L_{GK2} 
\equiv
%===revision===
%\textcolor{red}{
\sum_a
%}
%===
\int d^6 Z
\;
D_{a0} ({\bf Z}, t) f_{a0} ({\bf Z}, t) 
L_{GYa 2}  ({\bf Z}, t)
\end{equation}
where $f_{a0}$ is the zeroth-order part of the gyrocenter distribution 
function and $L_{GYa 2}$ is the second-order gyrocenter Lagrangian defined by
\begin{eqnarray}
\label{5-7}
 L_{GYa 2}  ({\bf Z}, t) 
% \equiv  - H_{GYa 2}  ({\bf Z}, t) 
& \equiv &
\frac{e_a}{c}  
%===revision===
%\textcolor{red}{
{\bf v}_{Ba}   
%}
%===
\cdot
\langle \widehat{\bf A} 
({\bf X} + \boldsymbol{\rho}_{a1}, t) 
\rangle_\xi
\nonumber \\
& & 
\mbox{}
\hspace*{-7mm}
- \frac{e_a^2}{2 m_a c^2} 
\langle
|\widehat{\bf A}
({\bf X} + \boldsymbol{\rho}_{a1}, t) |^2
\rangle_\xi
+ \frac{e_a^2}{2 B} 
\frac{\partial}{\partial \mu} 
\langle
(\widetilde{\psi}_a)^2
\rangle_\xi
.
\nonumber \\
& & 
\end{eqnarray}
We note here that $L_{GYa 1} + L_{GYa 2}$ corresponds to 
the opposite sign of $e_a \Psi_a$ defined by Eq.~(\ref{A-25}).  
The term 
$
(e_a^2/2 B) 
(\partial  
\langle
(\widetilde{\psi}_a)^2
\rangle_\xi
/ \partial \mu ) 
$
in Eq.~(\ref{5-7})
is a part of 
%===revision===
%\textcolor{red}{
$\frac{1}{2}e_a \langle 
\{ \widetilde{S}_a, \widetilde{\psi}_a \}
\rangle_\xi$ 
%}
%===
in Eq.~(\ref{A-25}), 
while the remaining part of 
%===revision===
%\textcolor{red}{
$\frac{1}{2}e_a \langle 
\{ \widetilde{S}_a, \widetilde{\psi}_a \}
\rangle_\xi$ 
%}
%===
is removed in $L_{GYa 2}$ because, when it is retained, 
its contribution  
to $L_{GK2}$ is of higher order in $\epsilon$
than that of the terms included in Eq.~(\ref{5-7}). 
As noted after Eq.~(\ref{A-26}) in Appendix~A, 
one of the second-order terms,  
$(e_a/c)  
%===revision===
%\textcolor{red}{
{\bf v}_{Ba}
%}
%===
\cdot 
\langle \widehat{\bf A} \rangle_\xi$, 
is often neglected 
in conventional studies although 
this term is kept here to accurately derive 
the gyrokinetic Amp\`{e}re's law later. 

The Lagrangian $L_F$ is defined by~\cite{Sugama2000}
\begin{eqnarray}
\label{5-8}
\label{LF}
L_F 
& \equiv & 
\frac{1}{8 \pi} 
\int_V d^3 x 
\left[
|{\bf E}_L ({\bf x}, t) |^2
- 
|{\bf B} ({\bf x}, t) + \widehat{\bf B} ({\bf x}, t)|^2
\right.
\nonumber \\ 
& & \mbox{}
\left. 
+ 
\frac{2}{c} \lambda ({\bf x}, t) 
\nabla \cdot \widehat{\bf A} ({\bf x}, t) 
\right]
. 
\end{eqnarray}
where the longitudinal (or irrotational) part 
${\bf E}_L$
of the electric field is 
written in terms of the electrostatic potential 
$\phi$ as
\begin{equation}
\label{5-9}
{\bf E}_L \equiv - \nabla \phi
,
\end{equation}
and $\lambda$ plays the role of the Lagrange undetermined multiplier to 
derive the Coulomb gauge condition, 
\begin{equation}
\label{5-10}
\nabla \cdot \widehat{\bf A} =0
,
\end{equation}
from $\delta I/ \delta \lambda = 0$ 
(or $\delta L_{GKF}/ \delta \lambda = \delta L_F/ \delta \lambda = 0$). 
Equation~(\ref{5-8}) is used for making the Darwin approximation to 
remove electromagnetic waves propagating at light speed.

From the condition that $\delta I = 0$ holds for the variation of ${\bf Z}_a (t)$ which is fixed at 
$t=t_1$, $t_2$, we can derive the gyrocenter motion equations for ${\bf Z}_a (t)$ and accordingly 
the gyrokinetic Vlasov equation for the distribution function $f_a$ which is constant along the 
gyrocenter phase-space orbit represented by ${\bf Z}_a (t)$. 
This is a variational derivation of the gyrokinetic Vlasov equation based on 
the Lagrangian picture of the gyrocenter phase-space 
motion.~\cite{Sugama2000} 
The resultant gyrokinetic Vlasov equation is given by removing the collision term 
from Eq.~(\ref{2-1}). 
In the Eulerian picture (or the Euler-Poincar\'{e} 
formulation),~\cite{Sugama2018,Sugama2021,Hirvijoki,Marsden,Cendra,Newcomb,Squire}
we use the expression in the last line of Eq.~(\ref{5-3}) and consider 
the variations of $f_a$ and $\dot{\bf Z}$ as functions of $({\bf Z}, t)$ 
to derive the gyrokinetic Vlasov equation from $\delta I = 0$. 
Effects of the collision term, if included, on 
the local energy and momentum balance equations can be 
clarified following the same procedure as shown in 
Refs.~\cite{Sugama2018,Sugama2021}. 

In the present case, Eq.~(\ref{5-6}) is used for the second-order Lagrangian 
to make the linear polarization-magnetization approximation,   
in which the deviation of $f_a$ from $f_{a0}$ does not enter
the polarization and magnetization terms proportional 
to $\phi$ and
$\widehat{\bf A}$ in the gyrokinetic Poisson and Amp\`{e}re equations 
as shown later.~\cite{Sugama2000} 
It also should be noted that 
in the gyrokinetic equation derived in this approximation,  
quadratic terms with respect to $\phi$ and
$\widehat{\bf A}$ are removed from 
the gyrocenter phase-space velocity $d{\bf Z}/dt$.

The gyrokinetic Poisson's equation is derived from the variational derivative of the action integral $I$ 
with respect to the electrostatic potential $\phi$. 
Since the time derivative of 
$\phi$ never appears in the Lagrangian density $L_{GKF}$, 
the above-mentioned condition can be replaced by 
$\delta L_{GKF}/\delta \phi=0$, which leads to
\begin{eqnarray}
\label{5-11}
\nabla \cdot {\bf E}_L
& = & 
4\pi \sum_a e_a 
\int d^6 Z
\; 
\delta^3 ({\bf X} + 
\boldsymbol{\rho}_{a1} - {\bf x} )
\biggl(
D_a 
f_a 
\nonumber \\ & & \mbox{}
+
D_{a0}
\frac{e_a \widetilde{\psi}_a}{B} 
\frac{\partial f_{a0}}{\partial \mu}
\biggr)
.
\end{eqnarray}

In ${\cal O}(e n_0)$ and ${\cal O}(\epsilon e n_0)$,  
the ensemble-averaged part of Eq.~(\ref{5-11}) gives 
the quasineutrality conditions,
\begin{equation}
\label{5-12}
0 = \sum_a e_a n_{a0} \equiv 
\sum_a e_a 
\int d^6 Z
\; D_{a0} f_{a0} 
\delta^3 ({\bf X} - {\bf x} )
,
\end{equation}
and 
\begin{equation}
\label{5-13}
0
= 
\sum_a e_a \langle n_{a1}^{(g)} \rangle_{\rm ens}
\equiv 
\sum_a e_a 
\int d^6 Z
\; D_{a0} \langle f_{a1} \rangle_{\rm ens}
\delta^3 ({\bf X} - {\bf x} )
,
\end{equation}
respectively. 
The fluctuation part of Eq.~(\ref{5-10}) is 
written as 
\begin{eqnarray}
\label{5-14}
\nabla \cdot \widehat{\bf E}_L
& = & 
4\pi \sum_a e_a 
\int d^6 Z
\; D_{a0} 
\delta^3 ({\bf X} + 
\boldsymbol{\rho}_{a1} - {\bf x} )
\biggl(
\widehat{f}_a 
\nonumber \\ & & \mbox{}
+
\frac{e_a \widetilde{\widehat{\psi}}_a}{B} 
\frac{\partial f_{a0}}{\partial \mu}
\biggr)
,
\end{eqnarray}
which is valid up to the lowest order, ${\cal O}(\epsilon e n_0)$. 
Here and hereafter, we do not consider 
the particle species dependence in using the ordering parameter   
$\epsilon \sim \rho_a/L$ and ${\cal O}(e_a n_{a0})$. 
Such dependence may occur due to large mass and charge 
differences although they should be treated using 
subsidiary parameters other than $\epsilon$. 
We can confirm that  
Eqs.~(\ref{5-11})--(\ref{5-14}) are consistent with 
the results derived from using Eqs.~(\ref{3-1}), 
(\ref{3-2}), and (\ref{A-11})  
for Poisson's equation 
$\nabla \cdot {\bf E}_L = 4\pi \sum_a e_a n_a^{(p)}$. 

The gyrokinetic  Amp\`{e}re's law is derived from 
the variational derivative of the action integral $I$ 
with respect to the fluctuation part $\widehat{\bf A}$ of the vector potential. 
Since the time derivative of 
$\widehat{\bf A}$ never appears in the Lagrangian density $L_{GKF}$,  
we can use  
$\delta L_{GKF}/\delta \widehat{\bf A}=0$ to obtain 
\begin{equation}
\label{5-15}
\nabla \times ( {\bf B} + \widehat{\bf B} ) 
=
\frac{4\pi}{c} {\bf j} -  \frac{1}{c} \nabla \lambda
, 
\end{equation}
where the electric current density is given by
\begin{eqnarray}
\label{5-16}
{\bf j} 
& = & 
\sum_a e_a 
\int d^6 Z
\; \delta^3 ({\bf X} + 
\boldsymbol{\rho}_{a1} - {\bf x} )
\Biggl[
D_a ({\bf Z}, t) f_a ({\bf Z}, t) {\bf v}_c 
\nonumber \\
& & 
\mbox{}
+
D_{a0} 
\left\{ 
 f_{a0} 
\biggl( 
%===revision===
%\textcolor{red}{
{\bf v}_{Ba}
%}
%=== 
- \frac{e_a}{m_a c} \widehat{\bf A} \biggr)
+
\frac{e_a \widetilde{\psi}_a}{B} 
\frac{\partial f_{a0}}{\partial \mu} {\bf v}_c 
\right\}
\Biggr]
.
\hspace*{8mm}
\end{eqnarray}
We see that the Eq.~(\ref{5-16}) agrees with the result shown in 
Eq.~(\ref{4B2-5}).
The longitudinal (or irrotational) part of Eq.~(\ref{5-15}) 
gives
\begin{equation}
\label{5-17}
\nabla \lambda
=
4\pi {\bf j}_L
.
\end{equation}
From the transverse (or solenoidal) part of Eq.~(\ref{5-15}), 
the gyrokinetic  Amp\`{e}re's law is written as  
\begin{equation}
\label{5-18}
\nabla \times ( {\bf B} + \widehat{\bf B} ) 
=
\frac{4\pi}{c} {\bf j}_T
. 
\end{equation}
In Eqs.~(\ref{5-17}) and (\ref{5-18}), ${\bf j}_L$ and ${\bf j}_T$ 
represent the longitudinal and transverse parts of 
${\bf j}$, respectively. 
It is noted here that 
an arbitrary vector field ${\bf a}$ is written as  
$
{\bf a} = {\bf a}_L + {\bf a}_T
$
where the longitudinal and transverse parts of ${\bf a}$ 
are given by
$
{\bf a}_L ({\bf x}) 
= - (4\pi)^{-1} \nabla \int d^3 x' \,
(\nabla' \cdot {\bf a} ({\bf x}') ) 
/ |{\bf x} - {\bf x}'|
$
and
$
{\bf a}_T ({\bf x}) 
= (4\pi)^{-1} \nabla \times 
( \nabla \times
\int d^3 x' \,
{\bf a} ({\bf x}') 
/ |{\bf x} - {\bf x}'| )
$, respectively.~\cite{Jackson}  

The ensemble-averaged part and the fluctuation part 
of Eq.~(\ref{5-18}) are written as 
\begin{equation}
\label{5-19}
\nabla \times {\bf B}  
=
\frac{4\pi}{c} \langle {\bf j} \rangle_{{\rm ens}\, T}
,
\end{equation}
and 
\begin{equation}
\label{5-20}
\nabla \times \widehat{\bf B}
=
\frac{4\pi}{c}
\widehat{\bf j}_T
, 
\end{equation}
respectively, 
where the ensemble-averaged part and fluctuation part of the 
current density is given by 
\begin{eqnarray} 
\label{5-21}
\hspace*{-5mm}
\langle {\bf j} \rangle_{\rm ens}
& = & 
\sum_a e_a 
\int d^6 Z
\; \delta^3 ({\bf X} + 
\boldsymbol{\rho}_{a1} - {\bf x} )
\Bigl[
D_{a0}  \langle f_{a1} \rangle_{\rm ens} {\bf v}_c 
\nonumber \\
& & 
\mbox{}
+ 
( D_{a0} + D_{a1} ) f_{a0} {\bf v}_c 
+
D_{a0} 
 f_{a0} 
{\bf v}_{da} 
\Bigr]
\nonumber \\ 
& = & 
\biggl\{
\sum_a e_a 
\int d^6 Z
\; \delta^3 ( {\bf X} - {\bf x} ) 
D_{a0} \langle f_{a1} \rangle_{\rm ens}
 U 
\nonumber \\ & & 
\mbox{} 
\hspace*{-2mm}
 + \frac{c}{B} 
\left( (P_\parallel)_0  - (P_\perp)_0 \right)
\left( {\bf b} \cdot \nabla \times {\bf b} \right)
\biggr\}
{\bf b}
\nonumber \\ & & 
\mbox{} 
\hspace*{-2mm}
+ \frac{c}{B} 
{\bf b}  \times 
\nabla \cdot 
\left\{ (P_\parallel)_0 {\bf b} {\bf b} 
+ (P_\perp)_0 
( {\bf I} - {\bf b} {\bf b} )
\right\}
,
\end{eqnarray}
and 
\begin{eqnarray}
\label{5-22}
\widehat{\bf j}
& = & 
 \sum_a e_a 
\int d^6 Z
\; \delta^3 ({\bf X} + 
\boldsymbol{\rho}_{a1} - {\bf x} )
D_{a0} 
\nonumber \\
& & 
\mbox{}
\cdot
\left( \widehat{f}_a  {\bf v}_c 
- f_{a0} 
 \frac{e_a}{m_a c} \widehat{\bf A}
+
\frac{e_a \widetilde{\widehat{\psi}}_a}{B} 
\frac{\partial f_{a0}}{\partial \mu} {\bf v}_c 
\right)
, 
\hspace*{5mm}
\end{eqnarray}
respectively. 
On the right-hand side of Eq.~(\ref{5-21}), 
$(P_\perp)_0 \equiv \sum_a (P_{a\perp})_0$ and 
$(P_\parallel)_0 \equiv \sum_a (P_{a\parallel})_0$ 
are used and the definitions of 
$(P_{a\perp})_0$ and $(P_{a\parallel})_0$ are found in  
Eqs.~(\ref{4B1-5}) and (\ref{4B1-7}), respectively. 
When $f_{a0}$ takes the form of the local Maxwellian distribution 
with no mean flow, 
we have
the isotropic equilibrium pressure $(P_\perp)_0 = (P_\parallel)_0 = P_0$. 
It should be noted that Eqs.~(\ref{5-16})
(\ref{5-21}), and (\ref{5-22}) are valid up to the lowest  
in $\epsilon$. 
In Appendix~D, using the WKB representation, 
the turbulent parts of Poisson and Amp\`{e}re equations 
in Eqs.~(\ref{5-14}) and (\ref{5-20}) are shown to 
agree with the results derived in earlier 
works.~\cite{Antonsen,CTB}

\section{CONCLUSIONS}

In this paper, 
effects of both equilibrium and gyroradius scale electromagnetic turbulence are 
included to derive expressions of polarization and magnetization in terms of the distribution function in the gyrocenter phase-space coordinates.
These expressions presented in Eqs.~(\ref{3-2}) and (\ref{3-7}) 
include infinite series expansion with respect to 
the gyroradius vector, which is defined in the gyrocenter coordinates by 
Eqs.~(\ref{A-11})--(\ref{A-17}),  
where effects of the turbulent fields are taken into account. 

%===revision===
%\textcolor{red}{
To the leading (or first) order in the normalized gyroradius parameter $\epsilon$, 
the polarization flow vanishes and 
the ensemble-averaged (or non-turbulent) part of the particle flow consists of 
the gyrocenter and magnetization flows, which agrees with the   
result called the magnetization law 
in the drift kinetics.~\cite{magnetization_law}
%}
%===
On the other hand, the leading-order turbulent part of the particle flow is 
given by the sum of the turbulent parts of the 
polarization, magnetization, and gyrocenter flows. 
%
%===revision===
%\textcolor{red}{
Thus, a practical extension of the  drift kinetic magnetization law  
is made to gyrokinetic systems with electromagnetic fluctuations and collisions.  
%}
%===
%
The compact expression of the particle flow, 
including both mean and 
turbulent parts, is given in Eq.(\ref{4B2-5}), 
which is valid to the leading order 
and useful for evaluating the total current density to 
self-consistently determine the magnetic field in full-$f$ global 
gyrokinetic 
simulations.~\cite{GTC,Idomura2017,XGC,Wang2009,GYSELA,ORB5,ELMFIRE,Gkeyll,Matsuoka} 

The effect of collisions appears as the classical transport 
in the second-order mean particle flow. 
In toroidally confined plasmas,
the first-order mean (or ensemble-averaged) particle flow is 
tangential to 
the magnetic surface, 
so that the mean particle transport 
flux across the magnetic surface is of 
the second-order and it is verified to contain classical, neoclassical, 
and turbulent transport processes which determine plasma particle 
confinement in a transport time scale.   

The Lagrangian is presented for variational derivation of 
the gyrokinetic Poisson and Amp\`{e}re equations, which 
properly include mean and turbulent parts. 
%
%===revision===
%\textcolor{red}{
It is shown that the diamagnetic current can be correctly included 
in the mean part of Amp\`{e}re's law derived from the variational principle 
using the Lagrangian, which retains 
the second-order term given 
by the inner product of the turbulent 
vector potential and the drift velocity consisting of 
the curvature drift and the $\nabla B$ drift. 
The resultant expressions of Amp\`{e}re's law [Eq.~(\ref{5-18})] and  
the current density [Eq.~(\ref{5-16})] are 
useful especially for the full-$f$ global electromagnetic gyrokinetic simulations 
to accurately treat high-beta plasmas.     
Properly taking account of the difference between the phase space coordinates in the 
classical gyrokinetic formulation and the modern formulation employed in the present work, 
the equivalence between descriptions of electromagnetic gyrokinetic turbulent fluctuations 
in the two formulations is clarified as shown in Appendices~C and D. 
The turbulent parts of the gyrokinetic Poisson and Amp\`{e}re equations 
in Eqs.~(\ref{5-11}) and (\ref{5-18}) 
are confirmed to agree with 
the results derived from the classical 
gyrokinetic formulation using the WKB representation in earlier works. 
%}
%===
Thus, these equations present a basic model for global 
full-$f$ gyrokinetic simulations which is also consistent with 
the local turbulence model used in flux-tube gyrokinetic 
simulations.~\cite{Dimits,GENE,GYRO,GKV,GKW} 
Based on the presented Lagrangian, 
local energy and momentum balance equations for the gyrokinetic 
system with electromagnetic turbulence and collisions 
can be derived following the same formulation 
as given by our previous work 
in the case of electrostatic turbulence.~\cite{Sugama2021} 
Details of the derivation will be reported in a future work.

\begin{acknowledgments}
This work is supported in part by 
the JSPS Grants-in-Aid for Scientific Research Grant No.~19H01879 
 and in part by the NIFS Collaborative Research Program NIFS20KNTT055. 
\end{acknowledgments}

\section*{AUTHOR DECLARATIONS}

\subsection*{Conflict of Interest}

The authors have no conflicts of interest to disclose.

\section*{DATA AVAILABILITY}
Data sharing is not applicable to this article as no new data were created or analyzed in this study.

\appendix

\section{GYROCENTER COORDINATES AND EQUATIONS OF MOTION}

We consider motion of a charged particle in a strong magnetic field. 
The particle mass and charge are denoted by $m_a$ and $e_a$, respectively, 
where the subscript $a$ represents the particle species. 
The magnetic field is assumed to consist of the 
background part ${\bf B} \equiv \nabla \times {\bf A}$ 
and the small fluctuation part 
$\widehat{\bf B} \equiv \nabla \times \widehat{\bf A}$. 
The particle's position and velocity 
are denoted by  
${\bf x}$ and ${\bf v}$, respectively. 
The velocity ${\bf v}$ 
is written by the sum of the parallel and 
perpendicular components as 
\begin{equation}
\label{A-1}
{\bf v} \equiv v_\parallel {\bf b} 
+ {\bf v}_\perp
,
\end{equation}
where the unit vector ${\bf b} \equiv {\bf B}/B$ in the direction 
parallel to the magnetic field is evaluated 
at the particle's position ${\bf x}$. 
Using a right-handed orthogonal triad of unit vectors 
$({\bf e}_1, {\bf e}_2, {\bf b})$ which 
are regarded as functions of $({\bf x}, t)$,  
we represent the perpendicular velocity as 
\begin{equation}
\label{A-2}
{\bf v}_\perp
\equiv 
- v_\perp 
( \sin \xi_0 \; {\bf e}_1 + \cos \xi_0 \; {\bf e}_2 )
,
\end{equation}
where $v_\perp \equiv |{\bf v}_\perp|$. 
We now define the particle phase-space coordinates 
${\bf z}$ by
\begin{equation}
\label{A-3}
{\bf z} \equiv 
({\bf x}, v_\parallel, \mu_0, \xi_0)
,
\end{equation}
where 
\begin{equation}
\label{A-4}
\mu_0 \equiv \frac{m_a v_\perp^2}{2B({\bf x}, t)}
.
\end{equation}

Using the Lie transformation technique, 
the gyrocenter phase-space coordinates, 
\begin{equation}
\label{A-5}
{\bf Z} \equiv 
({\bf X}, U, \mu, \xi)
,
\end{equation}
are obtained, such that the Lagrangian for the particle 
motion is transformed into a function which is independent 
of the gyrophase angle variable $\xi$, as shown later in 
Eq.~(\ref{A-20}). 
The relations of the gyrocenter coordinates 
${\bf Z} \equiv ({\bf X}, U, \mu, \xi)$
to the particle coordinates
${\bf z} \equiv ({\bf x}, v_\parallel, \mu_0, \xi_0)$
are given by
\begin{eqnarray}
\label{A-6}
{\bf X}
& = & 
{\bf x}
- 
\frac{v_\perp}{\Omega_a} {\bf a}
+
\frac{v_\perp}{\Omega_a^2} 
\biggl[ 
\biggl\{
v_\parallel ({\bf b} \cdot \nabla \times {\bf b})
- \frac{v_\perp}{2B} ({\bf a} \cdot \nabla B)
\biggr\} {\bf a}
\nonumber \\ 
& & \mbox{}
\hspace*{-5mm}
+ \biggl\{ 
2 v_\parallel 
({\bf b} \cdot \nabla {\bf b} \cdot {\bf c} )
+ \frac{v_\perp}{8}
({\bf c} \cdot \nabla {\bf b} \cdot {\bf c} 
- 5 {\bf a} \cdot \nabla {\bf b} \cdot {\bf a} ) 
\biggr\}
{\bf b}
\biggr]
\nonumber \\ & & \mbox{}
\hspace*{-5mm}
+ \frac{1}{B} \biggl[
\left(
\widehat{\bf A} + \frac{c}{e_a} \nabla \widetilde{S}_a
\right) \times {\bf b}
+ {\bf b} \int d\xi_0 \, \widetilde{\widehat{A}_\parallel}
\biggr]
,
\end{eqnarray}
\begin{eqnarray}
\label{A-7}
U
 & = & 
v_\parallel
- \frac{v_\perp}{\Omega_a} 
\biggl[
 v_\parallel ({\bf b} \cdot \nabla {\bf b} \cdot {\bf a} ) 
\nonumber \\ & & \mbox{}
+ \frac{v_\perp}{4}
(3 {\bf a} \cdot \nabla {\bf b} \cdot {\bf c} 
- {\bf c} \cdot \nabla {\bf b} \cdot {\bf a} ) 
\biggr]
+ \frac{e_a}{m_a c} \widehat{A}_\parallel
,
\hspace*{5mm}
\end{eqnarray}
\begin{eqnarray}
\label{A-8}
\mu
 & = & 
\frac{m_a v_\perp^2}{2B}
+
\frac{m_a v_\perp^2}{B \Omega_a}
\biggl[
\frac{v_\parallel^2}{v_\perp} 
({\bf b} \cdot \nabla {\bf b} \cdot {\bf a} )
\nonumber \\ & & \mbox{}
+ \frac{v_\parallel}{4}
(3 {\bf a} \cdot \nabla {\bf b} \cdot {\bf c} 
- {\bf c} \cdot \nabla {\bf b} \cdot {\bf a} ) 
+ \frac{v_\perp}{2B} ({\bf a} \cdot \nabla B)
\biggr]
\nonumber \\ & & \mbox{}
+ \frac{e_a}{B}
\left( \frac{1}{c} {\bf v}_\perp \cdot \widehat{\bf A}
+ \widetilde{\psi}_a \right)
,
\end{eqnarray}
and 
\begin{eqnarray}
\label{A-9}
\xi
 & = & 
\xi_0 
+
\frac{1}{\Omega_a}
\biggl[
\frac{v_\parallel^2}{v_\perp} 
({\bf b} \cdot \nabla {\bf b} \cdot {\bf c} )
+ \frac{v_\parallel}{4}
({\bf c} \cdot \nabla {\bf b} \cdot {\bf c} 
- {\bf a} \cdot \nabla {\bf b} \cdot {\bf a} ) 
\nonumber \\ & & \mbox{}
+ v_\perp
\left({\bf c} \cdot \frac{\nabla B}{B} 
- {\bf a} \cdot \nabla {\bf c} \cdot {\bf a} 
\right) 
\biggr]
- \frac{e_a}{m_a c}
\frac{\partial \widetilde{S}_a}{\partial \mu_0}
,
\end{eqnarray}
where $\Omega_a \equiv e_a  B({\bf x}, t)/ (m_a c)$, 
$v_\perp \equiv (2 \mu_0 B({\bf x}, t)/ m_a)^{1/2}$, 
${\bf c} \equiv {\bf v}_\perp/v_\perp$, 
${\bf a} \equiv {\bf b} \times {\bf c}$, 
$\widehat{A}_\parallel \equiv \widehat{\bf A} \cdot {\bf b}$, 
and the definitions of 
$\widetilde{\psi}_a$ and $\widetilde{S}_a$ are shown later 
in Eqs.~(\ref{A-18}) and (\ref{A-19}), respectively. 
Equation~(\ref{A-6}) for the gyrocenter position ${\bf X}$ is valid up 
to the second order in the normalized gyroradius parameter 
$\epsilon$, while 
Eqs.~(\ref{A-7})--(\ref{A-9}) are up to the first order. 
When there are no fluctuation fields, the formulas in 
Eqs.~(\ref{A-6})--(\ref{A-9}) agree with those given by 
Littlejohn,~\cite{Littlejohn1983} except that Eq.~(\ref{A-7}) 
is given here in a slightly different way,  
in order to remove the ${\cal O} (\epsilon)$ 
term of the Hamiltonian in Ref.~\cite{Littlejohn1983}. 
The same procedure as in Ref.~\cite{Sugama2000} is used to 
include the effects of the fluctuation fields in 
Eqs.~(\ref{A-6})--(\ref{A-9}).

We can inversely solve Eqs.~(\ref{A-6})--(\ref{A-9}) to 
represent the particle position vector ${\bf x}$ by 
the function of the gyrocenter coordinates ${\bf Z}$ as 
\begin{equation}
\label{A-10}
{\bf x}
=  {\bf X} + \boldsymbol{\rho}_a ({\bf Z}, t)
\end{equation}
where the gyroradius vector $\boldsymbol{\rho}_a ({\bf Z}, t)$ 
is expanded in $\epsilon$ as  
\begin{equation}
\label{A-11}
\boldsymbol{\rho}_a ({\bf Z}, t)
=  \boldsymbol{\rho}_{a1} ({\bf Z}, t) 
+ \boldsymbol{\rho}_{a2} ({\bf Z}, t) + \cdots
. 
\end{equation}
The lowest-order part of   
$\boldsymbol{\rho}_a$ is given by  
\begin{equation}
\label{A-12}
\boldsymbol{\rho}_{a1} ({\bf Z}, t)
\equiv 
\frac{{\bf b}({\bf X}, t) \times {\bf v}_c ({\bf Z}, t)}{\Omega_a({\bf X}, t)}
,
\end{equation}
where ${\bf v}_c$ is defined by  
\begin{equation}
\label{A-13}
{\bf v}_c
\equiv
U {\bf b} ({\bf X}, t)
-
W
[ \sin \xi \; 
{\bf e}_1 ({\bf X}, t) + \cos \xi \; 
{\bf e}_2 ({\bf X}, t) ]
,
\end{equation}
and 
\begin{equation}
\label{A-14}
W \equiv \left( \frac{2 \mu B ({\bf X}, t)}{m_a} \right)^{1/2}
.
\end{equation}
To the lowest order in $\epsilon$,
the particle velocity ${\bf v}$ and the gyroradius vector 
$\boldsymbol{\rho} \equiv {\bf x} - {\bf X}$ 
are represented by 
${\bf v}_c$ and $\boldsymbol{\rho}_{a1}$, respectively. 
The second-order part of   
$\boldsymbol{\rho}_a$ is written as 
\begin{equation}
\label{A-15}
\boldsymbol{\rho}_{a2} ({\bf Z}, t)
\equiv 
\langle \boldsymbol{\rho}_{a2} \rangle_{\rm ens}
+
\widehat{\boldsymbol{\rho}}_{a2}
,
\end{equation}
where 
the ensemble-average and fluctuation parts of 
$\boldsymbol{\rho}_{a2}$ are given by 
\begin{eqnarray}
\label{A-16}
\langle \boldsymbol{\rho}_{a2} \rangle_{\rm ens}
& \equiv & 
{\bf b}
\biggl[ 
- \frac{W^2}{8 \Omega_a^2} 
( 3 {\bf a} \cdot \nabla {\bf b} \cdot {\bf a} 
+ {\bf c} \cdot \nabla {\bf b} \cdot {\bf c} ) 
- \frac{2 U W}{\Omega_a^2} 
( {\bf b} \cdot \nabla {\bf b} \cdot {\bf c} ) 
\biggr]
\nonumber \\ 
& & \mbox{}
+ {\bf a}
\biggl[ 
\frac{c}{\Omega_a B} 
({\bf a} \cdot \langle {\bf E}_1 \rangle_{\rm ens} )
- \frac{W^2}{2 \Omega_a^2}
({\bf a} \cdot \nabla \ln B )
\nonumber \\ 
& & \mbox{}
+ \frac{U W}{4 \Omega_a^2}
( {\bf a} \cdot \nabla {\bf b} \cdot {\bf c} 
- 3  {\bf c} \cdot \nabla {\bf b} \cdot {\bf a} ) 
- \frac{U^2}{\Omega_a^2} 
( {\bf b} \cdot \nabla {\bf b} \cdot {\bf a} ) 
\biggr]
\nonumber \\ 
& & \mbox{}
+ {\bf c}
\biggl[ 
\frac{c}{\Omega_a B} 
({\bf c} \cdot \langle {\bf E}_1 \rangle_{\rm ens} )
- \frac{W^2}{\Omega_a^2}
({\bf c} \cdot \nabla \ln B )
\nonumber \\ 
& & \mbox{}
+ \frac{U W}{4 \Omega_a^2}
( {\bf a} \cdot \nabla {\bf b} \cdot {\bf a} 
- {\bf c} \cdot \nabla {\bf b} \cdot {\bf c} ) 
- \frac{U^2}{\Omega_a^2} 
( {\bf b} \cdot \nabla {\bf b} \cdot {\bf c} ) 
\biggr]
,
\nonumber \\ 
& & \mbox{}
\end{eqnarray}
and 
\begin{eqnarray}
\label{A-17}
& & \widehat{\boldsymbol{\rho}}_{a2}
=
\frac{\bf b}{B} \times \widehat{\bf A}
+
\{
{\bf X} + \boldsymbol{\rho}_{a1}, \widetilde{\widehat{S}}_a
\}
\nonumber \\ 
&  & 
=
- \frac{c}{B W} 
\left( 
\widehat{\phi} - \frac{U}{c} \widehat{A}_\parallel 
- \langle \widehat{\psi}_a \rangle_\xi 
\right)
{\bf a}
+ \frac{m_a c W}{B^2} \frac{\partial}{\partial \mu}
\left(
\int \widetilde{\widehat{\psi}}_a \, d\xi 
\right)
{\bf c}
\nonumber \\
&  & 
\mbox{}
- \frac{1}{B} 
\biggl[
\widehat{\bf A} + \frac{c}{\Omega_a} \nabla 
\left(
\int \widetilde{\widehat{\psi}}_a \, d\xi 
\right)
\biggr] \times {\bf b}
- \frac{1}{B} 
\left(
\int \widetilde{\widehat{A}}_\parallel \, d\xi 
\right)
 {\bf b}
,
\nonumber \\
&  & 
\end{eqnarray}
respectively. 
The definitions of $\langle \cdots \rangle_\xi$ and 
$\widetilde{\cdots}$ are given in 
Eqs.~(\ref{2-3}) and (\ref{2-4}), respectively, 
and   
$\widehat{\psi}_a \equiv  \psi_a 
- \langle \psi_a \rangle_{\rm ens}$ is 
the fluctuation part of $\psi_a$ 
which is defined in terms of 
the electrostatic potential 
$\phi$ 
and the fluctuation part 
$\widehat{\bf A}$ of the vector potential as 
\begin{equation}
\label{A-18}
\psi_a
\equiv
\phi ({\bf X} + \boldsymbol{\rho}_{a1}, t) 
- \frac{{\bf v}_c}{c} \cdot 
\widehat{\bf A}
({\bf X} + \boldsymbol{\rho}_{a1}, t) 
.
\end{equation}
Here, we also define
\begin{equation}
\label{A-19}
\widetilde{S}_a \equiv \frac{m_a c}{B}
\int \widetilde{\psi}_a d\xi
,
\end{equation}
where the integral constant is determined from the condition
$\langle \widetilde{S}_a  \rangle_\xi = 0$. 
We now note that 
$\widetilde{\widehat{\bf A}}$, 
$\widetilde{\psi}_a$, and $\widetilde{S}_a$ are 
defined above as 
functions of ${\bf Z}\equiv ({\bf X}, U, \mu, \xi)$ and $t$,  
although when they are substituted into the formulas for the coordinate 
transformation from ${\bf z}$ to ${\bf Z}$ 
[see Eqs.~(\ref{A-6})--(\ref{A-9})], the independent variables 
$({\bf X}, U, \mu, \xi)$ for the functions $\widetilde{\widehat{\bf A}}$, 
$\widetilde{\psi}_a$, and $\widetilde{S}_a$ should be replaced with 
$({\bf x} -  \boldsymbol{\rho}_{a1}({\bf z}, t), v_\parallel, 
\mu_0, \xi_0)$  to keep  
the validity of the formulas up to the orders described after Eq.~(\ref{A-9}).  
Here, the finite gyroradius $\boldsymbol{\rho}_{a1}$ cannot 
be neglected because fluctuations are considered to have 
${\cal O}(\rho_a)$ wavelengths 
in directions perpendicular to ${\bf B}$.

In the gyrocenter coordinates, 
the Lagrangian for the charged particle of motion    
is given by 
\begin{equation}
\label{A-20}
L_{GYa}  ({\bf Z}, \dot{\bf Z}, t)
  \equiv 
\frac{e_a}{c} {\bf A}_a^* 
\cdot \dot{\bf X}
+ \frac{m_a c}{e_a} \mu \; \dot{\xi}
- H_{GYa}  ({\bf Z}, t)
,
\end{equation}
where 
the modified vector potential ${\bf A}_a^*$ is 
defined by
\begin{equation}
\label{A-21}
{\bf A}_a^* 
 \equiv  
{\bf A} ({\bf X}, t)
+ \frac{m_a c}{e_a} U {\bf b} ({\bf X}, t)
- \frac{m_a c^2}{e_a^2} \mu 
{\bf W} ({\bf X}, t)
, 
\end{equation}
and 
\begin{equation}
\label{A-22}
{\bf W}
 \equiv  
\nabla {\bf e}_1 \cdot {\bf e}_2
+ \frac{1}{2} 
({\bf b} \cdot \nabla \times {\bf b}) {\bf b}
. 
\end{equation}
Here, the gyrocenter Hamiltonian $H_{GYa}$ is defined by
\begin{equation}
\label{A-23}
H_{GYa}  
 \equiv  
\frac{1}{2} m_a U^2 + \mu B 
+
e_a \Psi_a 
.
\end{equation}
The fluctuations are included in the Hamiltonian 
$H_{GYa}$ through the term $e_a \Psi_a$ defined by
%
%\begin{eqnarray}
%\label{A-24}
%e_a \Psi_a & \equiv  & 
% e_a \langle \psi_a \rangle_\xi
%- \frac{e_a}{c}  {\bf v}_{da}  \cdot 
%\langle \widehat{\bf A} \rangle_\xi
%+ \frac{e_a^2}{2 m_a c^2} 
%\langle |\widehat{\bf A}|^2 \rangle_\xi
%\nonumber \\ & & \mbox{}
%\hspace*{5mm}
%- \frac{e_a^2}{2 B} \frac{\partial}{\partial \mu} 
%\langle (\widetilde{\psi}_a)^2 \rangle_\xi
%,
%\end{eqnarray}
%
%
\begin{eqnarray}
\label{A-25}
e_a \Psi_a 
& \equiv  & 
 e_a \langle
\psi_a
\rangle_\xi
-
\frac{e_a}{c}  
%===revision===
%\textcolor{red}{
{\bf v}_{Ba}
%}
%===
  \cdot 
\langle \widehat{\bf A} \rangle_\xi
+ \frac{e_a^2}{2 m_a c^2} 
\langle
|\widehat{\bf A}|^2
\rangle_\xi
\nonumber \\ & & \mbox{}
\hspace*{5mm}
-
%===revision===
%\textcolor{red}{
\frac{e_a}{2} 
\langle
\{
\widetilde{S}_a, 
\widetilde{\psi}_a
\}
\rangle_\xi
%}
%===
,
\end{eqnarray}
where  $\{ \cdot, \cdot \}$ represents the Poisson bracket,  
defined by Eqs.~(29)--(33) in Ref.~\cite{Sugama2000}, and 
\begin{equation}
\label{A-26}
%===revision===
%\textcolor{red}{
{\bf v}_{Ba}
%}
%===
\equiv
\frac{c}{e_a B} {\bf b} \times 
\left( 
m_a U^2 {\bf b} \cdot \nabla {\bf b} 
+ \mu \nabla B 
\right)
\end{equation}
is the first-order drift velocity consisting of 
the curvature drift and the $\nabla B$ drift. 
On the right-hand side of Eq.~(\ref{A-25}), 
the first term is of ${\cal O}(\epsilon)$ and the others are 
of ${\cal O}(\epsilon^2)$. 
There the third and fourth terms are quadratic in the fluctuations,  
while the second term 
$- (e_a/c) 
%===revision===
%\textcolor{red}{
{\bf v}_{Ba}
%}
%===
 \cdot \langle \widehat{\bf A} \rangle_\xi$
is given by the product of the average drift velocity and 
the fluctuation vector potential. 
The latter term 
$- (e_a/c) 
%===revision===
%\textcolor{red}{
{\bf v}_{Ba}
%}
%===
 \cdot \langle \widehat{\bf A} \rangle_\xi$ 
is often neglected in conventional studies, although it is retained 
here for accuracy up to ${\cal O}(\epsilon^2)$.

The gyrocenter equations of motion are 
derived from the Euler-Lagrange equations using 
the gyrocenter Lagrangian in Eq.~(\ref{A-20}). 
Using the Hamiltonian in Eq.~(\ref{A-23}), they are given in the form, 
\begin{equation}
\label{A-27}
\frac{d{\bf Z}}{dt}
= 
\{ {\bf Z}, H_{GYa} \} + \{ {\bf Z}, {\bf X} \}
\cdot \frac{e_a}{c} \frac{\partial {\bf A}_a^*}{\partial t}
,
\end{equation}
which are rewritten as~\cite{Sugama2000} 
\begin{eqnarray}
\label{A-28}
\frac{d{\bf X}}{dt}
& = & 
\frac{1}{B^*_{a\parallel}}
\biggl[
\biggl( U + \frac{e_a}{m_a} 
\frac{\partial \Psi_a}{\partial U}
\bigg) {\bf B}_a^*
\nonumber \\
& & \mbox{}
\hspace*{5mm}
+
c {\bf b} \times \left( 
\frac{\mu}{e_a}\nabla B
+ \nabla \Psi_a
+ \frac{1}{c} \frac{\partial {\bf A}_a^*}{\partial t}
\right)
\biggl]
,
\hspace*{5mm}
\end{eqnarray}
\begin{equation}
\label{A-29}
\frac{dU}{dt}
=
-\frac{{\bf B}_a^*}{m_a B^*_{a\parallel}}
\cdot
\left(
\mu \nabla B
+
e_a \nabla \Psi_a
+ \frac{e_a}{c} \frac{\partial {\bf A}_a^*}{\partial t}
\right)
,
\end{equation}
\begin{equation}
\label{A-30}
\frac{d\mu}{dt}
=
0
,
\end{equation}
and 
\begin{eqnarray}
\label{A-31}
\frac{d\xi}{dt}
=
\Omega_a
+ 
{\bf W} \cdot \frac{d{\bf X}}{dt}
+
\frac{e_a^2}{m_a c} 
\frac{\partial \Psi_a}{\partial \mu}
,
\end{eqnarray}
where ${\bf B}_a^*$ and $B^*_{a\parallel}$ are defined in terms of  
${\bf A}_a^*$ in Eq.~(\ref{A-21}) as  
\begin{equation}
\label{A-32}
{\bf B}_a^*
\equiv
\nabla \times {\bf A}_a^*, 
\hspace*{3mm}
\mbox{and}
\hspace*{3mm}
B^*_{a\parallel}
\equiv
{\bf B}_a^* \cdot {\bf b}
,
\end{equation}
respectively. 
Since the gyrocenter Lagrangian $L_{GY}$ is 
independent of the gyrophase variable $\xi$, 
the time derivatives of the gyrocenter variables 
do not depend on $\xi$
and the magnetic moment 
$\mu = (e_a/m_a c)(\partial L_{GY}/\partial \dot{\xi})$ 
is conserved,  
as seen in Eqs.~(\ref{A-28})--(\ref{A-31}). 
The gyrocenter motion given by Eqs.~(\ref{A-28})--(\ref{A-31}) 
satisfies Liouville's theorem, which is expressed as
\begin{equation}
\label{A-33}
\frac{\partial D_a({\bf Z}, t)}{\partial t}
+ \frac{\partial}{\partial {\bf Z}}
\cdot \left(
 D_a ({\bf Z}, t)
\frac{d {\bf Z}}{d t}
\right)
=
0
, 
\end{equation}
where the Jacobian $D_a({\bf Z}, t)$ is given by
\begin{equation}
\label{A-34}
D_a({\bf Z}, t)
=
\frac{B^*_{a\parallel}}{m_a}
. 
\end{equation}

\section{EXPANSION OF $d{\bf X}/dt$ AND 
$d\boldsymbol{\rho}_a/dt$ IN $\epsilon$}

In this Appendix,  
$d{\bf X}/dt$ and $d\boldsymbol{\rho}_a/dt$ 
are expanded 
in the normalized gyroradius parameter $\epsilon$. 
To begin with, 
the zeroth-order gyrocenter velocity 
is parallel to the background magnetic field 
and given by 
\begin{equation}
\label{B-1}
\left( \frac{d{\bf X}}{dt} \right)_0 
=
U {\bf b} ({\bf X}, t)
,
\end{equation}
which contains no fluctuation part. 
The first-order gyrocenter velocity is written as 
\begin{equation}
\label{B-2}
\left( \frac{d{\bf X}}{dt} \right)_1 
=
\left\langle \left( 
\frac{d{\bf X}}{dt} \right)_1
\right\rangle_{\rm ens}
+  
\widehat{
 \left(
\frac{d{\bf X}}{dt}
\right)
}_1
,
\end{equation}
where the ensemble-averaged part and the fluctuation part 
are given by 
\begin{eqnarray}
\label{B-3}
\left\langle \left( 
\frac{d{\bf X}}{dt} \right)_1
\right\rangle_{\rm ens}
& = & 
\frac{c}{e_a B} {\bf b} \times 
\left( 
m_a U^2 {\bf b} \cdot \nabla {\bf b} 
+ \mu \nabla B 
+ e_a \nabla  \langle \phi_1 \rangle_{\rm ens}
\right)
\nonumber \\
& \equiv &
{\bf v}_{da}
,
\end{eqnarray}
and
\begin{eqnarray}
\label{B-4}
\widehat{
 \left(
\frac{d{\bf X}}{dt}
\right)
}_1
& = & 
-\frac{e_a}{m_a c}
\langle \widehat{A}_\parallel \rangle_\xi {\bf b}
+ \frac{c}{B} {\bf b} \times 
\nabla \langle \widehat{\psi}_a \rangle_\xi
\equiv 
\widehat{\bf v}_{ga}
,
\hspace*{5mm}
\end{eqnarray}
respectively. 
Regarding the second-order gyrocenter velocity, 
only its ensemble-averaged part is given here as 
\begin{eqnarray}
\label{B-5}
\left\langle \left( 
\frac{d{\bf X}}{dt} \right)_2
\right\rangle_{\rm ens}
& = & 
-\frac{U}{\Omega_a} \biggl[
 ({\bf b} \cdot \nabla \times {\bf b})
\, {\bf v}_{da}
+
\frac{\mu B}{m_a \Omega_a} (\nabla \times {\bf W})_\perp
\biggr]
\nonumber \\ & & \mbox{}
+ \frac{c}{B}
\left( - \nabla \langle \phi_2 \rangle_{\rm ens}
- \frac{1}{c} \frac{\partial {\bf A}}{\partial t}
\right) \times {\bf b}
\nonumber \\ 
& \equiv &
{\bf v}_{da2}
.
\end{eqnarray}

The zeroth-order part of 
$d\boldsymbol{\rho}_a/dt$ 
is given by the perpendicular component 
of the particle velocity as 
\begin{eqnarray}
\label{B-6}
\left( 
\frac{d\boldsymbol{\rho}_a}{dt} 
\right)_0
& = & 
\Omega_a 
\frac{\partial \boldsymbol{\rho}_{a1}}{\partial \xi}
=
({\bf v}_c)_\perp
\nonumber \\ 
& \equiv & 
-
\left( \frac{2 \mu B}{m_a} 
\right)^{1/2}
[ \sin \xi \; 
{\bf e}_1  + \cos \xi \; 
{\bf e}_2  ]
, 
\end{eqnarray}
The first-order part of 
$d\boldsymbol{\rho}_a/dt$ 
is written as 
\begin{equation}
\label{B-7}
\left( 
\frac{d\boldsymbol{\rho}_a}{dt} 
\right)_1
=
\left\langle
\left( 
\frac{d\boldsymbol{\rho}_a}{dt} 
\right)_1
\right\rangle_{\rm ens}
+
\widehat{\left( 
\frac{d\boldsymbol{\rho}_a}{dt} 
\right)_1
}
, 
\end{equation}
where
\begin{equation}
\label{B-8}
\left\langle
\left( 
\frac{d\boldsymbol{\rho}_a}{dt} 
\right)_1
\right\rangle_{\rm ens}
=
U {\bf b} \cdot 
\left( \nabla \boldsymbol{\rho}_{a1}
+ {\bf W}
\frac{\partial  \boldsymbol{\rho}_{a1}}{\partial \xi}
\right)
+
\Omega_a \frac{\partial \langle \boldsymbol{\rho}_{a2}
\rangle_{\rm ens}}{\partial \xi}
, 
\end{equation}
and 
\begin{equation}
\label{B-9}
\widehat{\left( 
\frac{d\boldsymbol{\rho}_a}{dt} 
\right)_1
}
=
-\frac{e_a}{m_a c}
\bigl(
 \widetilde{\widehat{A}_\parallel}  {\bf b}
+ \widehat{\bf A}_\perp
\bigr)
- \frac{c}{B} {\bf b} \times 
\nabla \langle \widehat{\psi}_a \rangle_\xi
+ \{ ({\bf v}_c)_\perp, 
\widetilde{\widehat{S}}_a \}
.
\end{equation}

The second-order ensemble-averaged part of 
$(c/e_a) {\bf M}_a$ is derived from 
Eqs.~(\ref{3-7}), (\ref{B-1}), (\ref{B-3}), (\ref{B-6}), 
and (\ref{B-8}) 
as 
\begin{eqnarray}
\label{B-10}
& & 
\left\langle
\left( 
\frac{c}{e_a} {\bf M}_a
\right)_2
\right\rangle_{\rm ens}
\nonumber \\ & & 
=
\frac{c}{e_a}
\int d^6 Z \; \delta^3 ( {\bf X} - {\bf x} )
( D_{a0} f_{a1} + D_{a1} f_{a0} ) ( - \mu {\bf b} )
\nonumber \\ & & 
\mbox{} \hspace*{3mm}
+ \frac{1}{\Omega_a} 
\biggl[
\int d^6 Z \; \delta^3 ( {\bf X} - {\bf x} ) 
D_{a0} f_{a0} U {\bf v}_{da}
\nonumber \\ & & 
\mbox{} \hspace*{3mm}
- 2 \frac{c}{e_a} \int d^6 Z \; \delta^3 ( {\bf X} - {\bf x} ) 
D_{a0} f_{a0} U \mu
{\bf b} \times ({\bf b}\cdot \nabla) {\bf b}
\nonumber \\ & & 
\mbox{} \hspace*{3mm}
+ \frac{c}{2 e_a} {\bf b} \times \nabla
\left(
\int d^6 Z \; \delta^3 ( {\bf X} - {\bf x} ) 
D_{a0} f_{a0} U \mu
\right)
\biggr]
,
\hspace*{5mm}
\end{eqnarray}
where $D_{a0}$ and $D_{a1}$ are given by 
\begin{equation}
\label{B-11}
D_{a0}
= 
\frac{B}{m_a}
, 
\hspace*{5mm}
D_{a1}
= 
\frac{c}{e_a} U {\bf b} \cdot ( \nabla \times {\bf b} )
.
\end{equation}

\section{ZEROTH AND FIRST-ORDER DISTRIBUTION FUNCTIONS}

We here consider the zeroth and first-order distribution functions 
in the normalized gyroradius parameter $\epsilon$, and present the 
kinetic equations satisfied by these distribution functions. 
As for the zeroth-order distribution function, 
Maxwellian and non-Maxwellian cases are treated.

\subsection{Case of Maxwellian zeroth-order distribution}

To the zeroth order in $\epsilon$, 
Eq.~(\ref{2-1}) is written as 
\begin{equation}
\label{C1-1}
 \dot{\bf Z}_0 \cdot 
\frac{\partial f_{a0}}{\partial  {\bf Z}}
=
\sum_b C_{ab}^{(p)} [ f_{a0}, f_{b0} ] 
,
\end{equation}
where $\dot{\bf Z}_0$ represents the zeroth-order part of 
$\dot{\bf Z} \equiv d {\bf Z}/d t$. 
The collision terms appear on the right-hand side of 
Eq.~(\ref{C1-1}) because 
the collision frequency is regarded here 
as of the same order as the transit 
frequency $\omega_{Ta}$. 

In Ref.~\cite{Hinton1976}, it is shown using Eq.~(\ref{C1-1}) and the property of the 
collision operator regarding the entropy production that,  
in the magnetic confinement system with nested toroidal magnetic surfaces, 
the collision term vanishes and $f_{a0}$ is the Maxwellian equilibrium 
$f_{aM}$ 
distribution function with no means flow, and satisfies
\begin{equation}
\label{C1-2}
{\bf b} \cdot \nabla
f_{aM} ( {\bf X}, {\cal E}_c, t )
= 0 
, 
\end{equation}
where 
${\cal E}_c$ represents the zeroth-order particle energy given by 
\begin{equation}
\label{C1-3}
{\cal E}_c 
= 
\frac{1}{2} m_a U^2 + \mu B + e_a \langle \phi_1 \rangle_{\rm ens}
. 
\end{equation}
It should be noted that, in Eq.~(\ref{C1-2}), 
$\nabla \equiv \partial /\partial {\bf X}$ 
acts on $f_{aM}$ with ${\cal E}_c$ fixed. 
Then we can write
\begin{eqnarray}
\label{C1-4}
f_{a0} 
& = &
f_{aM} ( {\bf X}, {\cal E}_c )
\nonumber \\
& = & 
n_{a0} 
\left(
\frac{m_a}{2\pi T_{a0}}
\right)^{3/2}
\exp 
\left( - 
\frac{{\cal E}_c - e_a \langle \phi_1 \rangle_{\rm ens}}{T_{a0}}
\right)
,
\hspace*{5mm}
\end{eqnarray}
where $n_{a0}$, $T_{a0}$ and $\langle \phi_1 \rangle_{\rm ens}$ 
need to be flux surface functions because of Eq.~(\ref{C1-2}). 

Next we find from Eq.~(\ref{2-1}) that 
the first-order ensemble-averaged gyrocenter distribution 
function $\langle f_{a1} \rangle_{\rm ens}$ satisfies 
\begin{eqnarray}
\label{C1-5}
& & 
\dot{\bf Z}_0 \cdot 
\frac{\partial \langle f_{a1} \rangle_{\rm ens}}{\partial  {\bf Z}}
+ \langle 
\dot{\bf Z}_1 
\rangle_{\rm ens}
\cdot 
\frac{\partial f_{aM}}{\partial  {\bf Z}}
\nonumber \\ 
& & =
 \sum_b (C_{ab}^{(p)})^L [\langle f_{a1} \rangle_{\rm ens}, 
\langle f_{b1} \rangle_{\rm ens} ]
\nonumber \\ 
& & \equiv 
\sum_b
%===revision===
%\textcolor{red}{
\langle
C_{ab}^{(p)} [\langle f_{a1} \rangle_{\rm ens}, f_{b0}]
+
C_{ab}^{(p)} [f_{a0}, \langle f_{b1} \rangle_{\rm ens}]
\rangle_\xi
%}
%===
,
\end{eqnarray}
where $(C_{ab}^{(p)})^L$ represents the linearized collision operator. 
Equation~(\ref{C1-5}) is 
the so-called linearized drift kinetic equation, 
which is used as a basic equation for the neoclassical transport 
theory.~\cite{Hinton1976,H&S,Helander} 

From the fluctuation part of Eq.~(\ref{C2-1}), 
the governing equation for the first-order fluctuation part of the
gyrocenter distribution function is obtained as 
\begin{equation}
\label{C1-6}
\frac{\partial}{\partial t} \widehat{f}_{a1}
+  
 \{ \widehat{f}_{a1}, {\cal E}_c \}
+
  \{ f_{aM} + \widehat{f}_{a1}, e \langle \widehat{\psi}_a \rangle_\xi \}
= 
%===revision===
%\textcolor{red}{
\langle
(C_{ab}^{(g)})^L [ \widehat{f}_{a1}, \widehat{f}_{b1} ]
\rangle_\xi
%}
%===
, 
\end{equation}
where effects of gyroradius scale 
perpendicular wavelengths of $\widehat{f}_{a1}$ 
are taken into account in defining the collision operator 
$(C_{ab}^{(g)})^L$ by 
\begin{eqnarray}
\label{C1-7}
& & 
(C_{ab}^{(g)})^L [\widehat{f}_{a1}, 
\widehat{f}_{b1} ]
\nonumber \\
& & \equiv \; 
e^{\boldsymbol{\rho}_{a1}\cdot \nabla} \; 
(C_{ab}^{(p)})^L [
e^{- \boldsymbol{\rho}_{a1}\cdot \nabla} \, \widehat{f}_{a1}, 
e^{- \boldsymbol{\rho}_{b1}\cdot \nabla} \, \widehat{f}_{b1} ]
.
\end{eqnarray}
Here, $\widehat{f}_{a1}$ is given by the sum of 
adiabatic and nonadiabatic parts as
\begin{equation}
\label{C1-8}
\widehat{f}_{a1}
=  
- \frac{e_a \langle \widehat{\psi}_a \rangle_\xi}{T_{a0}}
f_{aM} 
+ \widehat{h}_a
,
\end{equation}
which is substituted into Eq.~(\ref{C1-6}) to derive 
the equation for $\widehat{h}_a$,
\begin{eqnarray}
\label{C1-9}
& & 
\frac{\partial}{\partial t} \widehat{h}_a
+  
 \{ \widehat{h}_a, {\cal E}_c  + 
e \langle \widehat{\psi}_a \rangle_\xi \}
- \sum_b 
%===revision===
%\textcolor{red}{
\langle
(C_{ab}^{(g)})^L [ \widehat{h}_a, \widehat{h}_b ]
\rangle_\xi
%}
%===
\nonumber \\ 
&  & = 
 e_a \frac{\partial \langle \widehat{\psi}_a \rangle_\xi }{\partial t} 
\frac{f_{aM}}{T_{a0}}
-
\{ {\bf X}, e_a \langle \widehat{\psi}_a \rangle_\xi \}
\cdot 
\frac{f_{aM} ( {\bf X}, {\cal E}_c 
)}{\partial  {\bf X}}
.
\hspace*{10mm}
\end{eqnarray}

\subsection{Case of non-Maxwellian zeroth-order distribution}

In the zeroth-order in $\epsilon$, 
Eq.~(\ref{2-1}) gives 
\begin{equation}
\label{C2-1}
 \dot{\bf Z}_0 \cdot 
\frac{\partial f_{a0}}{\partial  {\bf Z}}
= 0,
\end{equation}
where the collision term is neglected by assuming the collision frequency 
to be sufficiently small. 
It is seen from Eq.~(\ref{C2-1}) 
that the zeroth-order distribution function 
$
f_0 =
f_0 ( {\bf X}, {\cal E}_c, \mu )
$
satisfies
\begin{equation}
\label{C2-2}
{\bf b} \cdot \nabla
f_{a0} ( {\bf X}, {\cal E}_c, \mu )
= 0 
,
\end{equation}
where ${\cal E}_c$ is defined in Eq.~(\ref{C1-3}) and 
$\nabla \equiv \partial /\partial {\bf X}$ 
acts on $f_{aM}$ with ${\cal E}_c$ fixed 
in the same way as in Eq.~(\ref{C1-2}).  

From the fluctuation part of Eq.~(\ref{2-1}), 
the governing equation for the first-order fluctuation part of the
gyrocenter distribution function is obtained as 
\begin{equation}
\label{C2-3}
\frac{\partial}{\partial t} \widehat{f}_{a1}
+  
 \{ \widehat{f}_{a1}, {\cal E}_c \}
+
  \{ f_{a0} + \widehat{f}_{a1}, e_a \langle \widehat{\psi}_a \rangle_\xi \}
= 
 \sum_b 
%===revision===
%\textcolor{red}{
\langle
(C_{ab}^{(g)})^L [\widehat{f}_{a1}, 
\widehat{f}_{b1} ]
\rangle_\xi
%}
%===
,
\end{equation}
where the collision term is retained for including collisional effects 
on gyrokinetic turbulence. 
Here, $\widehat{f}_{a1}$ is given by the sum of 
adiabatic and nonadiabatic parts as
\begin{equation}
\label{C2-4}
\widehat{f}_{a1}
=  
e_a \langle \widehat{\psi}_a \rangle_\xi
\frac{\partial f_{a0} ( {\bf X}, {\cal E}_c, 
\mu )}{\partial  {\cal E}_c}
+ \widehat{h}_a
\end{equation}
which is substituted into Eq.~(\ref{C2-3}) to derive 
the equation for $h_a$,
\begin{eqnarray}
\label{C2-5}
& & 
\frac{\partial}{\partial t} \widehat{h}_a
+  
 \{ \widehat{h}_a, {\cal E}_c  + 
e_a \langle \widehat{\psi}_a \rangle_\xi \}
-
 \sum_b 
%===revision===
%\textcolor{red}{
\langle
(C_{ab}^{(g)})^L [\widehat{h}_a, 
\widehat{h}_b ]
\rangle_\xi
%}
%===
\nonumber \\ 
& = & 
- e_a \frac{\partial \langle \widehat{\psi}_a \rangle_\xi
}{\partial t} 
\frac{\partial f_{a0} ( {\bf X}, {\cal E}_c, 
\mu )}{\partial  {\cal E}_c}
-
\{ {\bf X}, e_a \langle \widehat{\psi}_a \rangle_\xi \}
\cdot 
\frac{\partial f_{a0} ( {\bf X}, {\cal E}_c, 
\mu )}{\partial  {\bf X}}
.
\nonumber \\ & &
\end{eqnarray}
It is found that 
the nonlinear gyrokinetic equation in Ref.~\cite{F-C} can be reproduced 
from Eq.~(\ref{C2-5}) while neglecting the collision term and using 
the WKB representation described in Appendix~D.

Substituting Eq.~(\ref{C2-4}) 
into Eqs.~(\ref{5-14}) and (\ref{5-22}), 
the gyrokinetic Poisson and Amp\`{e}re equations are 
written as 
\begin{eqnarray}
\label{C2-6}
- \nabla^2 \widehat{\phi}
& = & 
4\pi \sum_a e_a 
\int d^3 X \, d{\cal E}_c \, d\mu \, d\xi
\sum_{\sigma=\pm 1}
\frac{B}{m_a^2 |U|}
\nonumber \\ & & \mbox{}
\times
\delta^3 ({\bf X} + 
\boldsymbol{\rho}_{a1} - {\bf x} )
\biggl[
e_a \widehat{\phi}
\frac{\partial f_{a0}}{\partial {\cal E}_c}
+
e_a \biggl(
\widehat{\phi} - \frac{U}{c} \widehat{A}_\parallel
\nonumber \\ & & \mbox{}
- \langle \widehat{\psi}_a \rangle_\xi 
\biggr)
\frac{1}{B}
\frac{\partial f_{a0}}{\partial \mu}
+ \widehat{h}_a 
\biggr]
,
\end{eqnarray}
and 
\begin{eqnarray}
\label{C2-7}
& & 
- \nabla^2 \widehat{\bf A}
=
\frac{4\pi}{c} \sum_a e_a 
\int d^3 X \, d{\cal E}_c \, d\mu \, d\xi
\sum_{\sigma=\pm 1}
\frac{B}{m_a^2 |U|}
\nonumber \\ & & \mbox{} \hspace*{5mm}
\times
\delta^3 ({\bf X} + 
\boldsymbol{\rho}_{a1} - {\bf x} )
\biggl[
U {\bf b}
\biggl\{
e_a \biggl(
\widehat{\phi} - \frac{U}{c} \widehat{A}_\parallel
- \langle \widehat{\psi}_a \rangle_\xi 
\biggr) 
\nonumber \\ & & \mbox{} \hspace*{5mm}
\times 
\frac{1}{B}
\frac{\partial f_{a0}}{\partial \mu}
+ \widehat{h}_a  
\biggr\}
+ ({\bf v}_c)_\perp \biggl\{
- \langle \widehat{\psi}_a \rangle_\xi 
\frac{1}{B}
\frac{\partial f_{a0}}{\partial \mu}
+ \widehat{h}_a  
\biggr\}
\biggr]
, 
\nonumber \\ & & \mbox{} 
\end{eqnarray}
respectively, where $\sigma \equiv U/|U|$ and 
$
|U| \equiv [ (2/m_a)( {\cal E}_c - \mu B 
- e_a \langle \phi_1 \rangle_{\rm ens}) ]^{1/2}
$
are used and the integration in 
${\cal E}_c$ and $\mu$ are done over the region defined by 
$0 \leq \mu B \leq  {\cal E}_c - e_a \langle \phi_1 \rangle_{\rm ens}$.

It is useful to consider a case in which the distribution function 
$f_a^{(p)}$ in the particle coordinates is used instead of 
the distribution function $f_a$ in the gyrocenter coordinates. 
These functions are related to each other by 
\begin{equation}
\label{C2-8}
f_a^{(p)}
({\bf x}, {\cal E}, \mu_0, \xi_0, t)
=
f_a
({\bf X}, {\cal E}_c, \mu, \xi, t)
,
\end{equation}
where ${\cal E}$ and ${\cal E}_c$ are used as independent variables 
instead of $v_\parallel$ and $U$, respectively. 
Here, following Ref.~\cite{Antonsen}, ${\cal E}$ is 
defined by 
\begin{equation}
\label{C2-9}
{\cal E} \equiv \frac{1}{2} m_a v^2 + e_a \Phi 
,
\end{equation}
where $\Phi$ is the equilibrium electrostatic 
potential and corresponds 
to $\langle \phi \rangle_{\rm ens}$ in our notation. 
The relation between ${\cal E}_c$ and ${\cal E}$ is 
written as 
\begin{equation}
\label{C2-10}
{\cal E}_c = {\cal E} + \varDelta {\cal E}
.
\end{equation}
Then, using Eqs.~(\ref{A-7}), (\ref{A-8}), (\ref{C1-3}), 
and (\ref{C2-9}),  
the fluctuation part $\varDelta \widehat{\cal E}$ of 
$\varDelta {\cal E}$ is obtained up to the leading order in 
$\epsilon$ as   
\begin{equation}
\label{C2-11}
\varDelta \widehat{\cal E} = 
e_a
\left(
\frac{1}{c}
{\bf v} \cdot \widehat{\bf A} 
+
\widetilde{\widehat{\psi}}_a
\right)
=
e_a
\left(
\widehat{\phi}
-
\langle \widehat{\psi}_a \rangle_\xi
\right)
.
\end{equation}
Equation~(\ref{A-8}) is rewritten as
\begin{equation}
\label{C2-12}
\mu = \mu_0 + \varDelta \mu
,
\end{equation}
and the fluctuation part $\varDelta \widehat{\mu}$ of 
$\varDelta \mu$ is given up to the leading order in 
$\epsilon$ as 
\begin{equation}
\label{C2-13}
\varDelta \widehat{\mu} = 
 \frac{e_a}{B} 
\left(
\frac{1}{c}
{\bf v}_\perp \cdot \widehat{\bf A} 
+
\widetilde{\widehat{\psi}}_a
\right)
=
 \frac{e_a}{B} 
\left(
\widehat{\phi}
- \frac{1}{c}
v_\parallel \widehat{A}_\parallel 
-
\langle \widehat{\psi}_a \rangle_\xi
\right)
.
\end{equation}
Noting that the zeroth-order parts of $f_a$ and $f_a^{(p)}$ 
are both given by the same function $f_{a0}$, 
and using Eqs.~(\ref{C2-4}), (\ref{C2-11}) and (\ref{C2-13}), 
the first-order fluctuation part $\widehat{f}_{a1}^{(p)}$ 
of $f_a^{(p)}$ is written as  
\begin{eqnarray}
\label{C2-14}
 & & 
\widehat{f}_{a1}^{(p)} ({\bf x}, {\cal E}, \mu_0, t )
\nonumber \\ 
&  & 
= \widehat{f}_{a1} 
({\bf x}-\boldsymbol{\rho}_{a1}, {\cal E}, \mu_0, t )
+
\left(
\varDelta \widehat{\cal E} 
\frac{\partial}{\partial {\cal E}}
+
\varDelta \widehat{\cal \mu} 
\frac{\partial}{\partial  \mu_0}
\right)
 f_{a0} ({\bf x}, {\cal E}, \mu_0, t ) 
\nonumber \\ 
&  & 
=
e_a \widehat{\phi} 
\frac{\partial f_{a0}}{\partial {\cal E}}
+ e_a 
\left(
\widehat{\phi} - \frac{v_\parallel}{c} \widehat{A}_\parallel
- \langle \widehat{\psi}_a \rangle_\xi
\right)
\frac{1}{B}
\frac{\partial f_{a0}}{\partial  \mu}
\nonumber \\ 
&  & \mbox{} \hspace*{6mm}
+ \widehat{h}_a ({\bf x}-\boldsymbol{\rho}_{a1}, {\cal E}, \mu_0, t )
,
\end{eqnarray}
We find from using Eq.~(\ref{C2-14})  
that 
Eqs.~(\ref{C2-6}) and (\ref{C2-7}) 
are rewritten in the well-known forms 
as 
$
- \nabla^2 \widehat{\phi}
= 
\sum_a e_a 
\int d^6 z' \, 
\delta^3 ({\bf x}' - {\bf x} ) 
\widehat{f}^{(p)} ({\bf z}')
$ 
and 
$
\nabla \times \widehat{\bf B}
= 
(4\pi/c)
\sum_a e_a 
\int d^6 z' \, 
\delta^3 ({\bf x}' - {\bf x} ) 
\widehat{f}^{(p)} ({\bf z}') {\bf v}'
$,
respectively.

\section{WKB REPRESENTATION}

Here, we consider any variable $Q$,  
the fluctuation part $\widehat{Q}$ of which has 
small wavelengths of the order of the gyroradius $\rho$ in directions  
perpendicular to the background magnetic field. 
Then we use the WKB (or ballooning) representation~\cite{Antonsen,CTB,F-C}  
for $\widehat{Q}$,
\begin{equation}
\label{D-1}
\widehat{Q} ({\bf x}, t)
= 
\sum_{{\bf k}_\perp} \widehat{Q}_{{\bf k}_\perp} ({\bf x}, t)
\exp [ i S_{{\bf k}_\perp} ({\bf x}, t) ]
,
\end{equation}
where $\widehat{Q}_{{\bf k}_\perp}({\bf x}, t)$ has 
the same gradient scale length $L$ as that of the equilibrium field,   
while the eikonal $S_{{\bf k}_\perp} ({\bf x}, t)$ represents 
the rapid variation with the wave number vector 
${\bf k}_\perp \equiv \nabla S_{{\bf k}_\perp} (\sim \rho^{-1})$ 
which satisfies ${\bf k}_\perp \cdot {\bf b} = 0$.

The first-order fluctuation part $\widehat{f}_{a1}^{(p)} ({\bf z}, t)$ 
of the distribution function in the particle coordinates is given by 
the WKB representation as
\begin{equation}
\label{D-2}
\widehat{f}_{a1}^{(p)} ({\bf z}, t)
= 
\sum_{{\bf k}_\perp} 
%===revision===
%\textcolor{red}{
\widehat{f}_{a1{\bf k}_\perp} ^{(p)}
({\bf z}, t)
%}
%===
\exp [ i S_{{\bf k}_\perp} ({\bf x}, t) ]
.
\end{equation}
The first-order fluctuation part $\widehat{f}_{a1} ({\bf Z}, t)$ 
of the gyrocenter distribution function and its 
nonadiabatic part $\widehat{h}_a ({\bf Z}, t)$ are given by 
the WKB representation as
\begin{equation}
\label{D-3}
\left[
\begin{array}{c}
\widehat{f}_{a1} ({\bf Z}, t)
\\
\widehat{h}_a ({\bf Z}, t)
\end{array}
\right]
= 
\sum_{{\bf k}_\perp} 
\left[
\begin{array}{c}
\widehat{f}_{a1{\bf k}_\perp} ({\bf Z}, t)
\\
\widehat{h}_{a{\bf k}_\perp} ({\bf Z}, t)
\end{array}
\right]
\exp [ i S_{{\bf k}_\perp} ({\bf X}, t) ]
,
\end{equation}
where the gyrocenter position vector ${\bf X}$ is used in 
the eikonal
$S_{{\bf k}_\perp} ({\bf X}, t)$ instead of the particle 
position vector ${\bf x}$. 
From Eqs.~(\ref{C2-4}) and (\ref{C2-14}), we have
\begin{equation}
\label{D-4}
\widehat{f}_{a1{\bf k}_\perp}
=  
e_a \langle \widehat{\psi}_a \rangle_{\xi{\bf k}_\perp}
\frac{\partial f_{a0} ( {\bf X}, {\cal E}_c, 
\mu )}{\partial  {\cal E}_c}
+ \widehat{h}_{a{\bf k}_\perp}
,
\end{equation}
and 
\begin{eqnarray}
\label{D-5}
& & 
\hspace*{-5mm}
\widehat{f}_{a1{\bf k}_\perp}^{(p)} 
=
e_a \widehat{\phi}_{{\bf k}_\perp}
\frac{\partial f_{a0}}{\partial {\cal E}_c}
+ e_a 
\biggl(
\widehat{\phi}_{{\bf k}_\perp} 
- \frac{U}{c}  \widehat{A}_{\parallel{\bf k}_\perp}
\nonumber \\ & & \mbox{}
- \langle \widehat{\psi}_a \rangle_{\xi{\bf k}_\perp} 
e^{-i {\bf k}_\perp \cdot \boldsymbol{\rho}_{a1} }
\biggr)
\frac{1}{B}
\frac{\partial f_{a0}}{\partial  \mu}
+  \widehat{h}_{a{\bf k}_\perp}
e^{-i {\bf k}_\perp \cdot \boldsymbol{\rho}_{a1} }
,
\end{eqnarray}
respectively, and 
Eq.~(\ref{C2-5}) is rewritten as 
\begin{eqnarray}
\label{D-6}
& & 
\left(
\frac{\partial}{\partial t} + U {\bf b} \cdot \nabla 
+ {\bf k}_\perp \cdot {\bf v}_{da} 
\right)
\widehat{h}_{a{\bf k}_\perp}
\nonumber \\
&  & \mbox{}
%===revision===
%\textcolor{red}{
- \sum_b
e^{i {\bf k}_\perp \cdot \boldsymbol{\rho}_{a1}} 
\langle
(C_{ab}^{(p)})^L [ \widehat{h}_a e^{-i {\bf k}_\perp \cdot \boldsymbol{\rho}_{a1}} ,  
\widehat{h}_b e^{-i {\bf k}_\perp \cdot \boldsymbol{\rho}_{b1}}]
\rangle_\xi
%}
%===
\nonumber \\ 
& = & 
- e_a 
\left( 
\frac{\partial f_{a0} }{\partial  {\cal E}_c}
\frac{\partial }{\partial t} 
+ i  \frac{c}{B} 
( {\bf b} \times {\bf k}_\perp ) \cdot 
\nabla f_{a0} 
\right)
\langle \widehat{\psi}_a \rangle_{\xi{\bf k}_\perp}
\nonumber \\ 
&  & \mbox{}
+  \frac{c}{B} 
\sum_{{\bf k}'_\perp + {\bf k}''_\perp = {\bf k}_\perp}
[ {\bf b} \cdot ( {\bf k}'_\perp \times {\bf k}''_\perp ) ]
\langle \widehat{\psi}_a \rangle_{\xi{\bf k}'_\perp}
\widehat{h}_{{\bf k}''_\perp}
,
\end{eqnarray}
where 
\begin{equation}
\label{D-7}
\langle \widehat{\psi}_a \rangle_{\xi{\bf k}_\perp}
=
J_0 \left( \frac{k_\perp W}{\Omega_a} \right)
\left(
\widehat{\phi}_{{\bf k}_\perp} 
- \frac{U}{c}  \widehat{A}_{\parallel{\bf k}_\perp}
\right)
+
J_1 \left( \frac{k_\perp W}{\Omega_a} \right)
\frac{W}{c}
\frac{\widehat{B}_{\parallel{\bf k}_\perp}}{k_\perp}
.
\end{equation}
Here, $J_0$ and $J_1$ are the first and second-order Bessel functions, 
respectively. 

In the WKB representation, 
the fluctuation part of the gyrokinetic Poisson's equation in 
Eq.~(\ref{C2-6}) and that of 
the gyrokinetic Amp\`{e}re's law in Eq.~(\ref{C2-7})
are given by
\begin{eqnarray}
\label{D-8}
& & 
k_\perp^2 \widehat{\phi}_{{\bf k}_\perp}
=
4\pi \sum_a e_a 
\int d{\cal E}_c \, d\mu \
\sum_{\sigma=\pm 1}
\frac{2\pi B}{m_a^2 |U|}
\biggl[
e_a \widehat{\phi}_{{\bf k}_\perp}
\frac{\partial f_{a0}}{\partial {\cal E}_c}
\nonumber \\ & & \mbox{}
\hspace*{6mm}
+
e_a \biggl(
\widehat{\phi}_{{\bf k}_\perp} 
- \frac{U}{c} \widehat{A}_{\parallel {\bf k}_\perp}
- J_0 (k_\perp W / \Omega_a) \langle \widehat{\psi}_a \rangle_{\xi{\bf k}_\perp}
\biggr)
\frac{1}{B}
\frac{\partial f_{a0}}{\partial \mu}
\nonumber \\ & & \mbox{}
\hspace*{6mm}
+ J_0 (k_\perp W / \Omega_a) \widehat{h}_{a{\bf k}_\perp} 
\biggr]
,
\end{eqnarray}
and
\begin{eqnarray}
\label{D-9}
& & 
k_\perp^2 \widehat{\bf A}_{{\bf k}_\perp}
=
\frac{4\pi}{c} \sum_a e_a 
\int d{\cal E}_c \, d\mu 
\sum_{\sigma=\pm 1}
\frac{2\pi B}{m_a^2 |U|}
\nonumber \\ & & \mbox{} 
\hspace*{1mm}
\times
\biggl[
U {\bf b}
\biggl\{
e_a \biggl(
\widehat{\phi}_{{\bf k}_\perp} 
- \frac{U}{c} \widehat{A}_{\parallel {\bf k}_\perp}
- J_0 (k_\perp W / \Omega_a) \langle \widehat{\psi}_a \rangle_{\xi{\bf k}_\perp}
\biggr) 
\nonumber \\ & & \mbox{} 
\hspace*{5mm}
\times 
\frac{1}{B}
\frac{\partial f_{a0}}{\partial \mu}
+ J_0 (k_\perp W / \Omega_a) \widehat{h}_{a{\bf k}_\perp} 
\biggr\}
\nonumber \\ & & \mbox{} 
\hspace*{1mm}
+ i \frac{{\bf b} \times {\bf k}_\perp}{k_\perp} W J_1 (k_\perp W / \Omega_a)\biggl\{
-  e_a \langle \widehat{\psi}_a \rangle_{\xi{\bf k}_\perp} 
\frac{1}{B}
\frac{\partial f_{a0}}{\partial \mu}
+  
\widehat{h}_{a{\bf k}_\perp} 
\biggr\}
\biggr]
,
\nonumber \\ & & \mbox{} 
\end{eqnarray}
respectively.
The component of Eq.~(\ref{D-9}) 
in the direction parallel to the background 
magnetic field is written as 
\begin{eqnarray}
\label{D-10}
& & k_\perp^2 \widehat{A}_{\parallel{\bf k}_\perp}
=
\frac{4\pi}{c} \sum_a e_a 
\int d{\cal E}_c \, d\mu 
\sum_{\sigma=\pm 1}
\frac{2\pi B}{m_a^2 |U|}
U 
\nonumber \\ & & \mbox{} 
\times
\biggl[
e_a \biggl(
\widehat{\phi}_{{\bf k}_\perp} 
- \frac{U}{c} \widehat{A}_{\parallel {\bf k}_\perp}
- J_0 (k_\perp W / \Omega_a) 
\langle \widehat{\psi}_a \rangle_{\xi{\bf k}_\perp}
\biggr) 
\frac{1}{B}
\frac{\partial f_{a0}}{\partial \mu}
\nonumber \\ & & \mbox{} \hspace*{5mm}
+ J_0 (k_\perp W / \Omega_a) \widehat{h}_{{\bf k}_\perp} 
\biggr]
,
\end{eqnarray}
where 
$\widehat{A}_{\parallel{\bf k}_\perp}
\equiv \widehat{\bf A}_{{\bf k}_\perp}
\cdot {\bf b}$.
Taking the inner product of Eq.~(\ref{D-9}) and 
$-i {\bf b} \times {\bf k}_\perp /k_\perp$ gives 
\begin{eqnarray}
\label{D-11}
&  & - k_\perp \widehat{B}_{\parallel{\bf k}_\perp}
=
\frac{4\pi}{c} \sum_a e_a 
\int d{\cal E}_c \, d\mu 
\sum_{\sigma=\pm 1}
\frac{2\pi B}{m_a^2 |U|}
\nonumber \\ & & \mbox{} 
\hspace*{5mm}
\times
W J_1 (k_\perp W / \Omega_a)
\biggl(
- e_a \langle \widehat{\psi}_a \rangle_{\xi{\bf k}_\perp}  
\frac{1}{B}
\frac{\partial f_{a0}}{\partial \mu}
+ \widehat{h}_{a{\bf k}_\perp} 
\biggr)
,
\nonumber \\ & & 
\end{eqnarray}
where 
$\widehat{B}_{\parallel{\bf k}_\perp} 
\equiv i ({\bf k}_\perp \times \widehat{\bf A}_{{\bf k}_\perp}) 
\cdot {\bf b}$. 
It is found from the inner product of Eq.~(\ref{D-9}) 
and ${\bf k}_\perp$ 
that the Coulomb gauge condition, 
${\bf k}_\perp \cdot \widehat{\bf A}_{{\bf k}_\perp}=0$, 
holds. 
Equations~(\ref{D-8}), (\ref{D-10}), and (\ref{D-11}) 
agree with the gyrokinetic Poisson and Amp\`{e}re equations derived in earlier works~\cite{Antonsen,CTB}  
using the WKB representation.

\end{document}